\documentclass[11pt]{article}
\usepackage{authblk}
\usepackage{cite}

\usepackage{graphicx,amsmath,amsfonts,mathtools}
\usepackage[margin=1.25in]{geometry}
\usepackage{soul,color}

\usepackage{tikz}
\usetikzlibrary{arrows}
\usetikzlibrary[patterns]

\newcommand{\vect}[1]{\boldsymbol{#1}}
\newcommand{\tensor}[1]{\boldsymbol{#1}}
\newcommand{\abs}[1]{\left| #1 \right|}

\newcommand{\normv}[1]{\| #1 \|}

\title{Connecting discrete and continuum dislocation mechanics: a non-singular spectral framework}
\author[1]{Nicolas Bertin}
\affil[1]{Department of Mechanical Engineering, Stanford University, Stanford, CA}
\affil[ ]{\textit {nbertin@stanford.edu}}
\date{\today}

\begin{document}

\maketitle
\begin{abstract}
In this paper, we present an improved framework of the spectral-based Discrete Dislocation Dynamics (DDD) approach introduced in \cite{Bertin15, Bertin18a}, that establishes a direct connection with the continuum Field Dislocation Mechanics (FDM) approach. To this end, an analytical method to convert a discrete dislocation network to its continuous dislocation density tensor representation is first developed. From there, the mechanical fields are evaluated using a FDM-based spectral framework, while submesh resolution elastic interactions are accounted for via the introduction of a rigorous stress splitting procedure that leverages properties of non-singular dislocation theories. The model results in a computationally efficient approach for DDD simulations that enables the use of elastic anisotropy and heterogeneities, while being fully compatible with recently developed subcycling time-integrators. As an example, the model is used to perform a work-hardening simulation, and potential applications that take advantage of the full-field nature of the method are explored, such as informing FDM models, and efficiently calculating virtual diffraction patterns.
\end{abstract}

\section{Introduction}

Dislocation mechanics has emerged as a fundamental field of research towards understanding the plastic behavior of metals during deformation. Over the years, several numerical methods have been developed to investigate dislocation-mediated plasticity at small scales. Among them, Field Dislocation Mechanics (FDM) was recently introduced to model the mechanical fields arising from a continuous distribution of crystalline defects (e.g. dislocations, disclinations) in a medium subjected to mechanical loadings \cite{Acharya01, Berbenni14}. Based on a continuum formulation, the FDM relies on the dislocation density tensor to obtain a full description of the mechanical state of the distorted body, for which finite element \cite{Roy05} and spectral-based implementations have been proposed for homogeneous \cite{Brenner14, Berbenni14} and heterogeneous \cite{Djaka17} media.

However, as any continuum descriptions, the FDM model does not treat the microstructure as an explicit collection of individual defects, and therefore may lack to resolve critical features of dislocation-mediated plasticity such as dislocation short-range interactions and core reactions. In that sense, the evolution laws in FDM are entailed to be governed by the collisionless dynamics of homogenized density fields, and a global connection between internal stresses and the resulting velocity fields is still lacking. As a result, the applications of the FDM have been limited thus far to the investigation of simple 1D configurations \cite{XZhang15}, static problems in 2D \cite{Brenner14, Berbenni14, Djaka17}, or by relying on phenomenological approaches \cite{Acharya06} or prescribing constant velocity fields \cite{Djaka15}.

On the other hand, front-tracking methods such as Discrete Dislocation Dynamics (DDD), which simulate the motion and interactions of large ensemble of dislocations, have demonstrated their ability to generate large-scale discrete dislocation structures that capture salient physical features of the plastic deformation \cite{Arsenlis07, Devincre08}, while unraveling the critical role of dislocation core reactions \cite{Bulatov06, SillsBertinAghaeiCai}. However, the incorporation of a more detailed physical description in these models comes at the price of a high computational cost. Specifically, dislocations in DDD models are represented as a set of inter-connected segments whose mutual elastic interactions must be explicitly calculated, making the approach a $N$-body simulation. To alleviate the computational cost of such simulations, substantial efforts have been undertaken in the recent years. Among them, an efficient full-field approach, referred to as DDD-FFT \cite{Bertin15, Bertin18a}, was recently developed in an attempt to transform the computational nature of such simulations. Based on the eigenstrain formalism proposed in \cite{Lemarchand01, Vattre14}, the core idea of the method consists in computing the mechanical fields directly from the plastic strain distribution associated with dislocation motion using a spectral method. In addition to being computationally efficient, the spectral-based formulation inherently allows for the use of anisotropic elasticity, and can be conveniently extended to heterogeneous elasticity \cite{Bertin18a}, thereby paving the way to performing discrete simulations of plasticity in polycrystalline materials. Nonetheless, the original DDD-FFT framework \cite{Bertin15, Bertin18a} suffers from some limitations that will be pointed out in the following. The objective of this work is to propose an improvement of the DDD spectral approach that addresses some of these limitations. 

In this paper, we propose a reformulation of the DDD-FFT framework that introduces a direct connection between DDD and FDM approaches. The core idea of the model relies on a framework in which interaction stresses (and forces) are decomposed into two components in a rigorous manner by taking advantage of the non-singular dislocation theory \cite{Cai06}. Specifically, the discrete dislocation structure is first converted into a non-singular continuous dislocation tensor field representation on a discrete grid. From there, the mechanical fields are computed using a spectral FDM framework and the resulting stress is used to evaluate the long-range component of the elastic forces driving the motion of the dislocation structure. Supplementary local stress contributions are then added between close dislocation segments to capture short-range interactions that are not resolved by the spectral grid. From there, the motion of dislocation segments can be time-integrated and topological operations can be performed on the discrete dislocation structure following a traditional DDD approach.

In essence, the core idea of this novel formulation remains identical to the original DDD-FFT approach, and the reformulation mainly consists in establishing a self-consistent non-singular framework while employing a different spectral solver. Importantly, we point out that the resort to the dislocation density tensor offers several advantages compared to the original framework:

\begin{enumerate}
\item Unlike the original history-dependent formulation based on the plastic strain distribution, the dislocation tensor only depends on the current dislocation configuration. As such, the calculation of the mechanical state at any given step needs no prior knowledge of the past dislocation structure. In addition, this solution eliminates the numerical errors associated with the plastic strain regularization likely to accumulate when performing dynamics simulations.
\item The approach provides a unique stress field solution thanks to the orthogonal field decomposition used in FDM that separates compatible and incompatible parts. As a result, the translation tensors introduced in \cite{Bertin15} to ensure coincidence between short and long range contributions can be eliminated.
\item The formulation provides a convenient framework that can incorporate the non-singular theory of dislocation \cite{Cai06} in a consistent manner.
\end{enumerate}

The last item in the above list is certainly the most important improvement proposed in this work compared to the original DDD-FFT approach. Practically, the non-singular spectral framework provides the following benefits:

\begin{enumerate}
\item The framework relies on an accurate description of the dislocation core that is consistent with the majority of recent implementations of DDD simulations.
\item The formulation allows to introduce a rigorous splitting between short-range and long-range force contributions, thereby rendering the treatment of dislocation interactions fully independent of the grid resolution.
\item The non-singular formulation provides a numerically adjustable force splitting radius, allowing the coupling with the efficient subcycling time-integration scheme recently developed in \cite{Sills16}.
\item The solution offers a way to eliminate the spurious Gibbs oscillations by providing a numerical, yet physically-based spreading function for discrete dislocation fields.
\end{enumerate}

The rest of this paper is organized as follows. In section \ref{sec:alphagrid}, an analytical method is presented to convert an arbitrary discrete dislocation network into its continuous dislocation tensor field representation. The non-singular spectral framework to evaluate the mechanical fields arising from the dislocation structure is then presented in section \ref{sec:alphaspectral}, after which the validity of the model is assessed in section \ref{sec:validation} by examining a static and a dynamics case. In section \ref{sec:applications}, the application of the model to DDD simulations is discussed, while additional benefits provided by the model are illustrated through two different applications, namely the potential ability of the model to inform FDM models and the efficient computation of virtual X-ray diffraction patterns. Finally, a discussion and a conclusion are given in sections \ref{sec:discussion} and \ref{sec:conclusion}.

\section{Discrete to continuous dislocation representation} \label{sec:alphagrid}

In this section, an analytical approach to convert a discrete dislocation structure to its continuum density representation is presented.

Consider a continuous medium $\Omega$ with displacement field $\vect{u}$ in which dislocations are present. The Burgers vector $\vect{b}$ associated with a dislocation line $L$ is defined as:

\begin{equation} \label{eq:Burgers}
b_i = \oint_{C} du_i = \oint_{C} u_{i,j}dx_j
\end{equation}

\noindent where $C$ denotes any circuit irreducibly enclosing the dislocation line. Let $\tensor{U} = \tensor{U}^e +\tensor{U}^p = \mathbf{grad}\,\tensor{u}$ denote the distortion field (i.e. the displacement gradient) defined as the sum of an elastic $\tensor{U}^e$ and a plastic $\tensor{U}^p$ part. In the small strain setting and applying Stokes' theorem to Eq.~\eqref{eq:Burgers}, it can be shown that the Burgers vector can be expressed as:

\begin{equation} \label{eq:Burgers2}
b_i = \oint_{C} U^e_{ij}dx_j = \int_{S} e_{klj} U^e_{ij,l} dS_k \equiv \int_{S} \alpha_{ik} dS_k
\end{equation}

\noindent where $S$ denotes the surface spanning contour $C$, $e_{ijk}$ is the permutation tensor, and $\tensor{\alpha} = \mathbf{curl}\,\tensor{U}^e = -\mathbf{curl}\,\tensor{U}^p$ is the dislocation density tensor field, referred to as Nye's tensor \cite{Nye53}.

From the definition of the Burgers vector in Eq.~\eqref{eq:Burgers2}, the dislocation density tensor field associated with a dislocation line segment $L^s$ with Burgers vector $\vect{b}^s$ is commonly given by:

\begin{equation} \label{eq:alphadef}
\alpha_{ij}^s(\vect{x}) = \int_{L^s} \delta(\vect{x}-\vect{x'}) b_i^s t_j^s(\vect{x'}) \; dL(\vect{x'})
\end{equation}

\noindent where $\delta(\vect{x}-\vect{x'})$ is the three-dimensional Dirac delta function vanishing everywhere except on line segment $L^s$ spanned with coordinate $\vect{x'}$, and $\vect{t}^s(\vect{x'})$ is the unit tangent of dislocation line $L^s$ at position $\vect{x'}$.

When considering a discrete dislocation network composed of $N_{\rm seg}$ inter-connected dislocation segments, the total dislocation density field is given by:

\begin{equation} \label{eq:alphatot}
\alpha_{ij}(\vect{x}) = \sum_s^{N_{\rm seg}}\alpha_{ij}^s(\vect{x})
\end{equation}

\noindent where the sum is carried over all dislocation segments $s$ present in the volume.

Nye's tensor field in Eq.~\eqref{eq:alphatot} fully describes the dislocation content in volume $\Omega$; it contains information on both the induced distortion and the orientation of all dislocation lines. Clearly, Eq.~\eqref{eq:alphatot} provides a continuous description of the total dislocation content. In the rest of this section, we present an efficient method to numerically evaluate the dislocation tensor on a discrete grid.

Consider a discrete grid $\Omega^d$ mapping the simulation volume $\Omega$. In general, the position and line contour of dislocations do not coincide with grid vertices; it is therefore necessary to weight the continuous distribution in Eq~\eqref{eq:alphatot} on every grid point $\{\vect{x}^d\}$. From a general perspective, the total dislocation tensor $\tensor{\alpha}(\vect{x}^d)$ at discrete grid point $\vect{x}^d$ can be determined by evaluating the following integral:

\begin{equation}
\alpha_{ij}(\vect{x}^d) = \int_{\Omega}  S^d(\vect{x}^d-\vect{x}) \alpha_{ij}(\vect{x}) \; d\Omega 
\label{eq:alphadiscrete}
\end{equation}

\noindent where $S^d(\vect{x}^d-\vect{x})$ denotes a weighting function associated with vertex $d$ of the discrete grid, and the integral is carried out over all $\vect{x} \in \Omega$. Using the definition of the total dislocation tensor in Eq.~\eqref{eq:alphatot} and the property of the delta function in Eq.~\eqref{eq:alphadef}, the construction of the discrete dislocation field is given by:

\begin{equation}
\alpha_{ij}(\vect{x}^d) = \sum_s^{N_{\rm seg}} b_i^s \int_{L^s}  S^d(\vect{x}^d-\vect{x}) t_j^s(\vect{x}) \; dL(\vect{x}) 
\label{eq:alphadiscrete2}
\end{equation}

\noindent where the integral is carried along each line segment $L^s$. The main difficulty associated with Eq.~\eqref{eq:alphadiscrete2} generally lies in the evaluation of the line integral, which may turn into a computational burden when performed numerically. In the following, we present a method to evaluate Eq.~\eqref{eq:alphadiscrete2} analytically for straight dislocation segments.

Let us consider the discrete dislocation tensor $\tensor{\alpha}^s(\vect{x}^d)$ associated with a single straight dislocation segment $s$. Since the segment is straight, its tangent $\vect{t}^s = \vect{t}^s(\vect{x})$ is constant over the line $L^s$, and the evaluation of $\tensor{\alpha}^s(\vect{x}^d)$ reduces to:

\begin{equation}
\alpha_{ij}^s(\vect{x}^d) = b_i^s t_j^s \int_{L^s} S(\vect{x}^d-\vect{x}) \; dL(\vect{x}) \equiv b_i^s t_j^s \, I(\vect{x}^d,L^s)
\label{eq:alphadiscreteSeg}
\end{equation}

\noindent When segment line $L^s$ is delimited by end points at positions $\vect{x}^a$ and $\vect{x}^b$, the integral $I(\vect{x}^d,L^s)$ in Eq.~\eqref{eq:alphadiscreteSeg} rewrites:

\begin{equation}
I(\vect{x}^d,L^s) = \int_{\vect{x}^a}^{\vect{x}^b} S(\vect{x}^d-\vect{x}) \; dL(\vect{x})
\label{eq:IntSeg}
\end{equation}

\noindent The evaluation of Eq.~\eqref{eq:IntSeg} is essentially tied to the choice of the weight function $S(\vect{x}^d-\vect{x})$. In this work, we propose to use the Cloud-In-Cell (CIC) weighting scheme \cite{Birdsall69}, whose three-dimensional expression is given by:

\begin{equation}
 S(\vect{x}^d-\vect{x}) =  
  \begin{dcases} 
   \prod_{i=1}^3 \left( 1 - \frac{\abs{x^d_i-x_i}}{H_i} \right) & \text{when } |x_i^d-x_i| < H_i,\; \forall i \in \{1,2,3\} \\
   0       & \text{otherwise }
  \end{dcases}
\label{eq:CIC3D}
\end{equation}

\noindent where $H_i$ denotes the grid spacing in each spatial direction $i = \{1,2,3\}$. The choice of the CIC weighting function in Eq.~\eqref{eq:CIC3D} is primarily motivated by the fact that an analytical coordinate-independent expression for Eq.~\eqref{eq:IntSeg} can be conveniently obtained by appropriately parametrizing segment $L^s$, as detailed in Appendix \ref{app:CIC}. 

Note that with this approach, the scalar dislocation density field $\rho(\vect{x}^d)$ is directly obtained as:

\begin{align}
\rho(\vect{x}^d) &= \frac{1}{V_d}\sum_s^{N_{\rm seg}} \int_{L^s} S^d(\vect{x}^d-\vect{x}) \; dL(\vect{x}) = \frac{1}{V_d}\sum_s^{N_{\rm seg}} I(\vect{x}^d,L^s) \nonumber\\
\rho^{\rm tot} &= \frac{V_d}{\Omega} \sum_d^{N_d} \rho(\vect{x}^d)
\label{eq:discreterho}
\end{align}

\noindent where $V_d = H_1 \times H_2 \times H_3$ is the volume associated with each voxel, $N_d$ is the total number of grid points, and $\rho^{\rm tot}$ denotes the total dislocation density in volume $\Omega$. The dislocation density per slip system can be simply obtained by distinguishing the slip systems associated with each dislocation segment in Eq.~\eqref{eq:discreterho}.

\begin{figure}[t]
  \begin{minipage}[b]{0.32\linewidth}
  \begin{center}
    \includegraphics[scale=0.33]{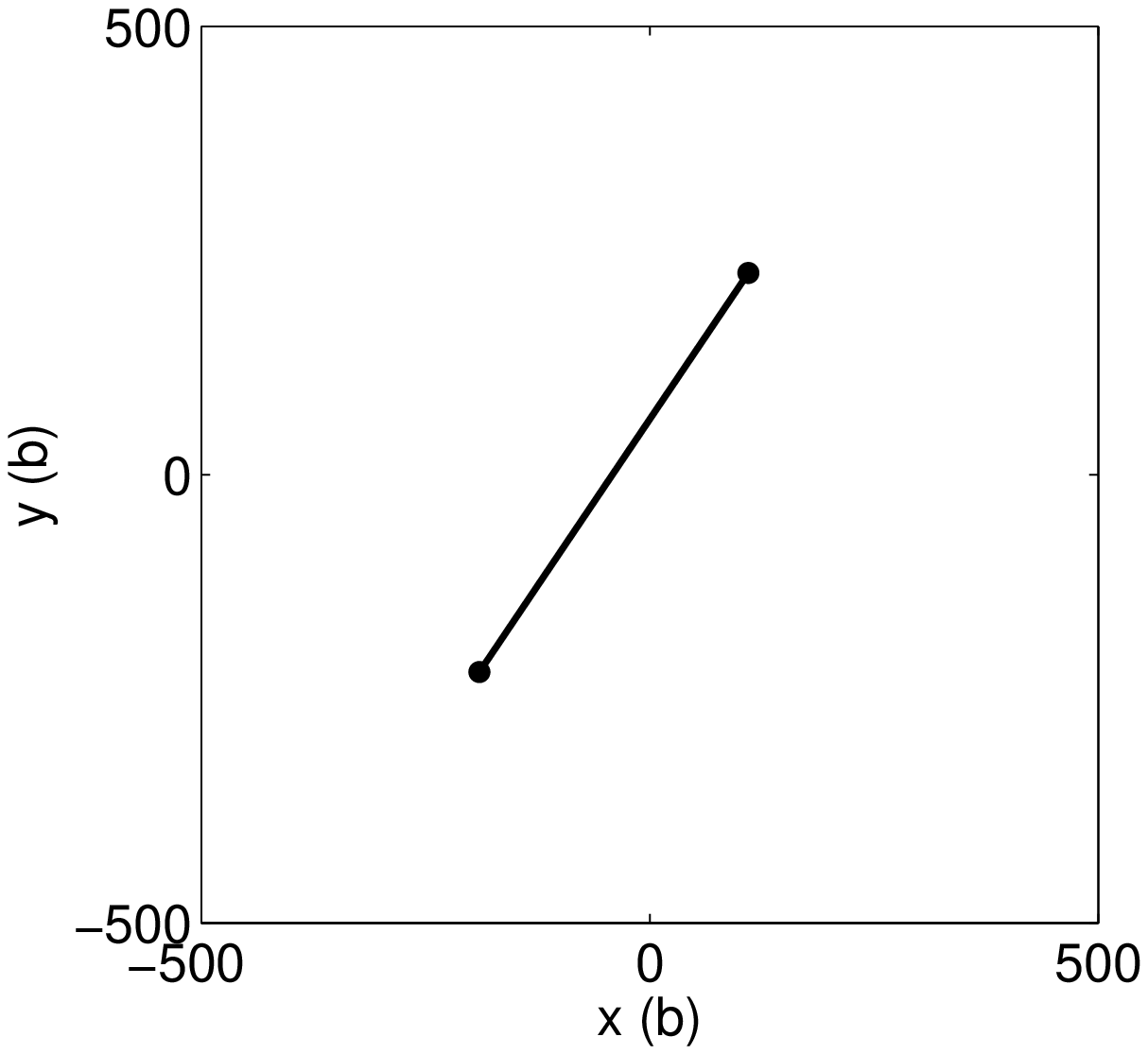} \\ (a)
  \end{center}
  \end{minipage}
  \begin{minipage}[b]{0.32\linewidth}
  \begin{center}
    \includegraphics[scale=0.33]{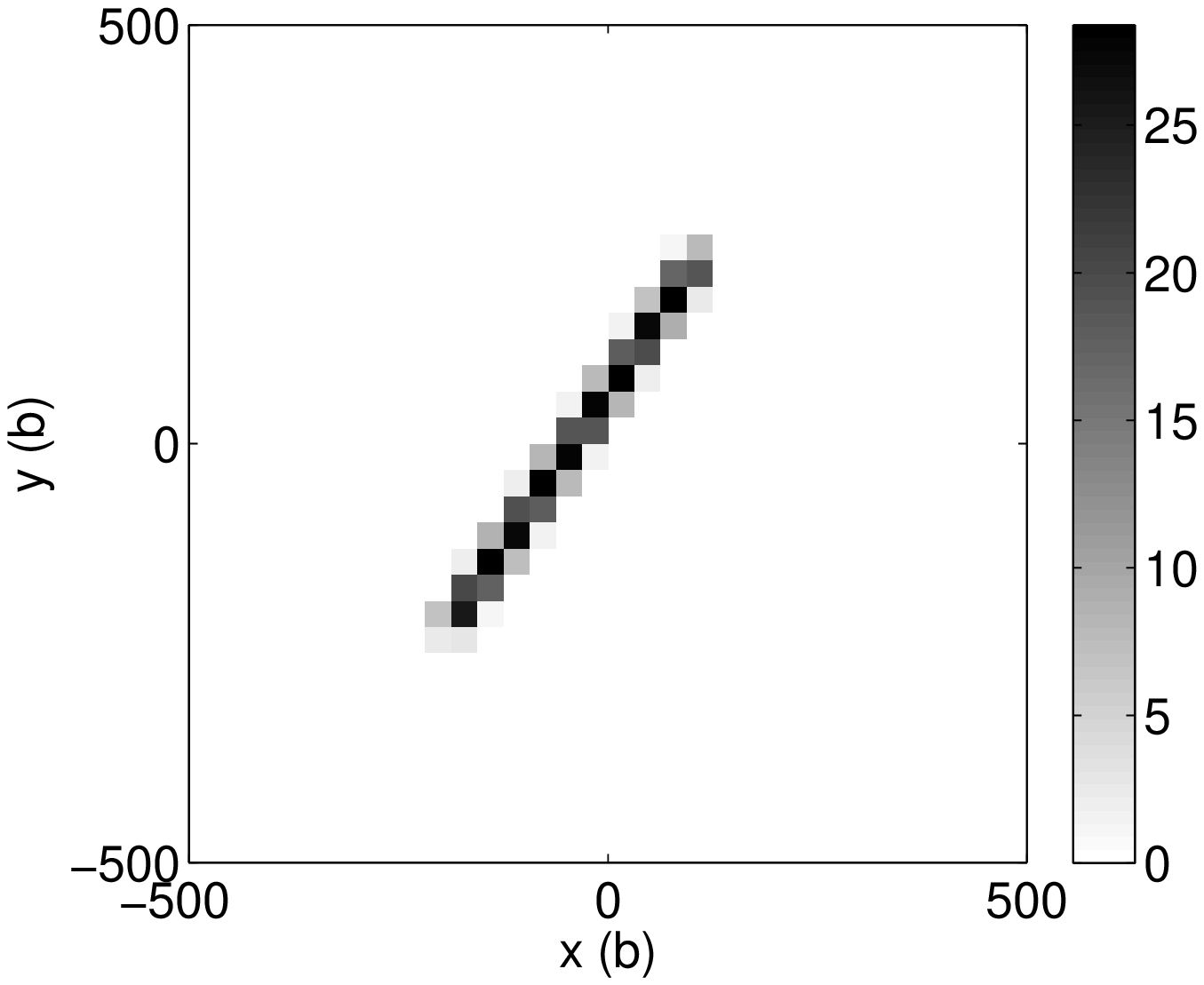} \\ (b)
  \end{center}
  \end{minipage}
  \begin{minipage}[b]{0.32\linewidth}
  \begin{center}
    \includegraphics[scale=0.33]{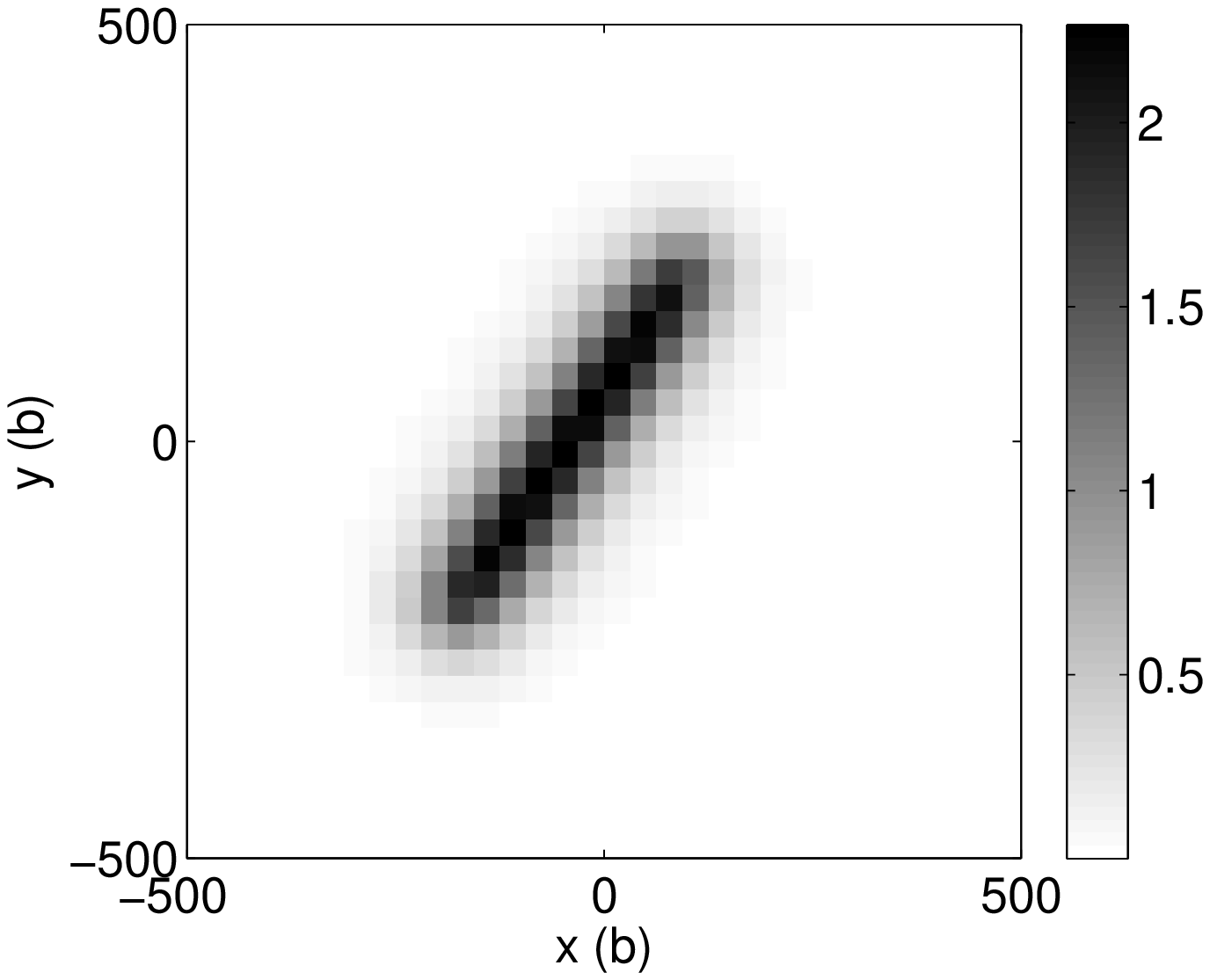} \\ (c)
  \end{center}
  \end{minipage}
  \caption{Example of the discrete to continuous density field conversion. (a) Discrete dislocation segment $L^s$ lying on a $xy$-slice. (b) Continuous representation of the density field $\rho(\vect{x}^d)$ calculated on a grid of resolution $N_d = 32^3$ voxels using the CIC scheme and the analytical approach for the line integral $I(\vect{x}^d,L^s)$. (c) Non-singular density field $\rho^{\rm ns}(\vect{x}^d,a_H) = \rho(\vect{x}^d) \ast \omega(\vect{x}^d,a_H)$ (see Eq. \eqref{eq:alphaconv}) for a spreading radius $a_H=100b$. The convolution is computed using a FFT.}
  \label{fig:segmentconv}
\end{figure}

An example of the discrete to continuous approach presented in this section is illustrated in Fig.~\ref{fig:segmentconv}. For the sake of clarity, the scalar dislocation density field $\rho(\vect{x}^d)$ given in Eq.~\eqref{eq:discreterho} is calculated for a single dislocation segment on a grid of resolution $N_d = 32^3$ voxels, as shown in Fig.~\ref{fig:segmentconv}(b). To achieve so, the CIC scheme and the analytical formulation for the line integral $I(\vect{x}^d,L^s)$ are used. The dislocation density tensor $\alpha_{ij}(\vect{x}^d)$ is obtained in a similar fashion by including the Burgers vector and line direction information.

\section{Spectral-based approach for periodic dislocation fields} \label{sec:alphaspectral}

\subsection{Non-singular stress field decomposition} \label{sec:stressdecomp}

The analytical approach presented in section \ref{sec:alphagrid} provides an efficient way to compute, on a discrete grid, the dislocation tensor field $\tensor{\alpha}(\vect{x}^d)$ associated with an arbitrary three-dimensional dislocation network, from which one seeks to evaluate the corresponding stress field. However, the discrete nature of the dislocation tensor field $\tensor{\alpha}(\vect{x}^d)$ renders the accuracy of the solution fields necessarily dependent on the grid resolution. Specifically, considering the fact that the grid spacing in numerical calculations is typically orders of magnitude larger than the length-scale associated with dislocations (e.g. Burgers vector magnitude), it necessarily follows that submesh short-range elastic interactions between dislocation segments cannot be resolved by a grid-based approach alone.

To address this issue and recover accurate solution fields that are fully independent of the grid resolution while enabling submesh resolution features to be captured, we propose a framework that allows for a rigorous splitting between short-range and long-range field contributions. The core idea of this approach relies on tailoring specific mathematical properties associated with non-singular theories of dislocation.

Specifically, consider the isotropic Burgers vector distribution $\omega(\vect{x},a)$ introduced by Cai and co-workers in \cite{Cai06}:

\begin{equation} \label{eq:omega}
\omega(\vect{x},a) = \frac{15}{8\pi a^3 \left[ (r/a)^2 +1\right]^{7/2}}, \quad r = \normv{\vect{x}}
\end{equation}

\noindent where $a$ is a parameter introduced to define the width of the dislocation core radius. In principle however, $a$ can be treated as an adjustable numerical parameter. With this in mind, two different but yet consistent length-scales can be introduced, each being associated with a given core radius. Specifically, let denote $a_0 \sim b$ the physical core radius used to describe stress fields at dislocation cores (where $b$ is the Burgers vector magnitude), and $a_H$ a spreading radius associated with the grid spacing $H_i$. Taking advantage of the property that the non-singular stress field arising from distribution~\eqref{eq:omega} converges at large distance $r$ for any two distinct values of $a$ (e.g. see Appendix \ref{app:splittingerror}), the total stress field associated with physical core radius $a_0$ can be conveniently decomposed as follows:

\begin{equation} \label{eq:stresstot}
\sigma_{ij}^{\rm tot}(\vect{x},a_0) = \sigma_{ij}^{\rm long}(\vect{x},a_H) + \sigma_{ij}^{\rm short}(\vect{x},a_0,a_H)
\end{equation}

\noindent Essentially, Eq.~\eqref{eq:stresstot} introduces a mathematically rigorous stress splitting procedure that involves two separate contributions:

\begin{enumerate}
\item A long-range stress contribution $\tensor{\sigma}^{\rm long}(\vect{x},a_H)$ directly obtained from the dislocation tensor field convolved with the non-singular kernel $\omega(\vect{x},a_H)$, following the approach detailed in \S\ref{sec:longrange}
\item A short-range correction $\sigma_{ij}^{\rm short}(\vect{x},a_0,a_H)$ obtained from existing analytical expressions developed to compute individual segment pairs interaction forces, as presented in \S\ref{sec:shortrange}
\end{enumerate}

Note that in the case of anisotropic elasticity, the isotropic core distribution in Eq.~\eqref{eq:omega} can still be used, provided that the same distribution is employed to remove the core singularity in the anisotropic formulation used to calculate short-range contributions. Alternatively, dedicated anisotropic core distributions such as that recently developed in \cite{Po2017} could also be used due to their similar parametric nature.

\subsection{Long-range stress contribution} \label{sec:longrange}

In this section, the spectral approach to evaluate the long-range stress field arising from the dislocation density tensor is presented.

Consistently with the splitting scheme introduced in Eq.~\eqref{eq:stresstot}, we first note that the long-range stress contribution $\tensor{\sigma}^{\rm long}(\vect{x},a_H)$ must possess the non-singular character of the distribution in Eq.~\eqref{eq:omega} with spreading radius $a_H$. For any arbitrary dislocation structure (represented as a density tensor), this is simply achieved by spreading the Burgers vector along all dislocation lines using the non-singular kernel $\omega(\vect{x},a_H)$, i.e. by calculating a non-singular dislocation tensor $\tensor{\alpha}^{\rm ns}(\vect{x},a_H)$ as follows:

\begin{equation} \label{eq:alphaconv}
\alpha_{ij}^{\rm ns}(\vect{x},a_H) = \alpha_{ij}(\vect{x}) \ast \omega(\vect{x},a_H)
\end{equation}

\noindent where $\ast$ denotes the convolution operator in the real-space, and $\tensor{\alpha}(\vect{x})$ is the discrete density tensor field computed using the method presented in section \ref{sec:alphagrid}. Numerically, applying the core spreading in Eq.~\eqref{eq:alphaconv} requires that the non-singular distribution width spreads over multiple voxels, i.e. $a_H$ must be chosen to be larger than the grid spacing. When dealing with a periodic dislocation structure, $\tensor{\alpha}^{\rm ns}(\vect{x},a_H)$ can be efficiently computed using Fast Fourier Transforms (FFT), as will be detailed later in this section. An example of the non-singular density tensor obtained for a dislocation segment is presented in Fig.~\ref{fig:segmentconv}(c).

From there, the framework to determine the long-range stress field contribution directly follows the spectral implementation proposed in \cite{Brenner14} in the context of FDM. Consider a periodic dislocation density tensor field $\tensor{\alpha}^{\rm ns}$ in a continuous body $\Omega$. Using the Stokes-Helmholtz orthogonal decomposition, it can be shown that there exist a vector field $\vect{z}$ and a second-order tensor $\tensor{\chi}$ such that the elastic distortion $\tensor{U}^e$ can be expressed as \cite{Brenner14}:

\begin{equation}
\tensor{U}^e = \mathbf{grad}\,\vect{z} + \tensor{\chi}
\end{equation}

\noindent where $\vect{z}$ is unique, and $\tensor{\chi}$ is unique up to a constant tensor and satisfies:

\begin{align}
\mathbf{curl}\,\tensor{\chi} &= \tensor{\alpha}^{\rm ns} \nonumber\\
\mathbf{div}\,\tensor{\chi} &= \tensor{0}
\end{align}

\noindent With this, the mechanical state in the periodic volume can be determined by solving the following fundamental problem of mechanics:

\begin{align}
&\tensor{\sigma}^{\rm long} = \tensor{C}:\tensor{U}^e = \tensor{C}:\left( \mathbf{grad}\,\vect{z} + \tensor{\chi} \right) \label{eq:sigmarel} \\
&\mathbf{div}\,\tensor{\sigma}^{\rm long} = \tensor{0} \label{eq:sigmaeq} \\
&\left\langle \tensor{\sigma}^{\rm long} \right\rangle = \tensor{\Sigma}
\end{align}

\noindent where $\tensor{C}$ is the fourth-order stiffness tensor of the medium, possibly anisotropic, and $\tensor{\Sigma}$ is the macroscopic imposed stress.

In the general case of heterogeneous elasticity, the mechanical fields can be solved for following the polarization scheme developed by Moulinec and Suquet \cite{Moulinec98}. In index notation, the constitutive relation subjected to the mechanical equilibrium in Eqs. \eqref{eq:sigmarel}-\eqref{eq:sigmaeq} can be written as:

\begin{align}
&C^0_{ijkl} z_{k,lj}(\vect{x}) + \tau_{ij,j}(\vect{x}) = 0 \label{eq:eqrealspacehetero} \\
&\tau_{ij}(\vect{x}) = \left( C_{ijkl}(\vect{x}) - C^0_{ijkl} \right) z_{k,l}(\vect{x}) + C_{ijkl}(\vect{x}) \chi_{kl}(\vect{x}) \label{eq:polarhetero}
\end{align}

\noindent where $\tensor{\tau}$ is the so-called polarization tensor, and $\tensor{C}^0$ is a fictitious reference medium introduced to account for spatial variations in the elastic stiffness. Note that in the case of an homogeneous medium, the stiffness tensor $\tensor{C}(\vect{x}) = \tensor{C}^0$ is constant throughout the simulation domain $\Omega$ such that the polarization tensor in Eq.~\eqref{eq:polarhetero} reduces to:

\begin{equation} \label{eq:polarhomo}
\tau_{ij}(\vect{x}) = C^0_{ijkl} \chi_{kl}(\vect{x})
\end{equation}

\noindent When all fields are periodic in the three spatial directions, the equilibrium equation \eqref{eq:eqrealspacehetero} can be generally expressed in the Fourier space as:

\begin{equation} \label{eq:eqFourierspace}
C^0_{ijkl} k_l k_j \widehat{z}_k(\vect{k}) - \mathit{i} k_j\widehat{\tau}_{ij}(\vect{k}) = 0
\end{equation}

\noindent where $\vect{k}$ denotes the frequency vector in the Fourier space, $\mathit{i}$ is the imaginary number, and $\widehat{\vect{z}}$ and $\widehat{\vect{\tau}}$ are the Fourier transforms of fields $\vect{z}$ and $\vect{\tau}$, respectively. Denoting $K_{ik} = C^0_{ijkl} k_l k_j$, Eq. \eqref{eq:eqFourierspace} can be rewritten as:

\begin{align} \label{eq:zFourier}
&\widehat{z}_{i}(\vect{k}) = \mathit{i} k_j \widehat{L}_{ik}(\vect{k}) \widehat{\tau}_{kj}(\vect{k}) \nonumber\\
&\widehat{L}_{ik}(\vect{k}) = K_{ki}^{-1} = \left[ C^0_{kjil} k_l k_j \right]^{-1} \quad \forall \vect{k} \neq \vect{0}
\end{align}

\noindent where $\widehat{\tensor{L}}(\vect{k})$ is the periodic Green's function in the Fourier space. Using Eq. \eqref{eq:zFourier}, the solution strain field, given by the symmetric part of $\mathbf{grad}\,\vect{z}$, is obtained as:

\begin{align}
&\widehat{\varepsilon}_{ij}(\vect{k}) = -\widehat{\Gamma}_{ijkl}(\vect{k}) \widehat{\tau}_{kl}(\vect{k}) \label{eq:strainFourier} \\
&\widehat{\Gamma}_{ijkl}(\vect{k}) = \frac{1}{2} \left( k_l k_j \widehat{L}_{ik}(\vect{k}) + k_l k_i \widehat{L}_{jk}(\vect{k})  \right) \quad \forall \vect{k} \neq \vect{0} \label{eq:GreenFourier}
\end{align}

\noindent where $\widehat{\tensor{\Gamma}}(\vect{k})$ is the modified Green's operator. When considering heterogeneous elasticity, Eq.~\eqref{eq:strainFourier} provides an implicit expression for the solution strain field that needs to be solved using iterative schemes (e.g. see \cite{Bertin18a}). Combining Eq. \eqref{eq:strainFourier} and the constitutive relation in Eq. \eqref{eq:sigmarel}, the stress field in the Fourier space is obtained as:

\begin{align} \label{eq:stressFourier}
&\widehat{\sigma}_{ij}^{\rm long}(\vect{k}) = \widehat{ \left[ C:\varepsilon \right]}_{ij}(\vect{k}) + \widehat{ \left[ C:\chi \right] }_{ij}(\vect{k}) \quad \forall \vect{k} \neq \vect{0} \nonumber\\
&\widehat{\sigma}_{ij}^{\rm long}(\vect{0}) = \Sigma_{ij}
\end{align}

\noindent where the Fourier transform of tensor field $\tensor{\chi}$ is given by construction as \cite{Brenner14}:

\begin{align} \label{eq:chiFourier}
&\widehat{\chi}_{ij}(\vect{k}) = \mathit{i} \frac{e_{jkl} k_l \widehat{\alpha}_{ik}^{\rm ns}(\vect{k},a_H) }{\normv{\vect{k}}^2} \quad \forall \vect{k} \neq \vect{0} \nonumber\\
&\widehat{\chi}_{ij}(\vect{0}) = 0
\end{align}

\noindent In Eq.~\eqref{eq:chiFourier}, $\widehat{\tensor{\alpha}}^{\rm ns}$ is the Fourier transform of the non-singular dislocation density tensor, whose expression in the Fourier space is readily given from Eq.~\eqref{eq:alphaconv} as:

\begin{equation} \label{eq:alphaconvFourier}
\widehat{\alpha}_{ij}^{\rm ns}(\vect{k},a_H) = \widehat{\alpha}_{ij}(\vect{k}) \widehat{\omega}(\vect{k},a_H)
\end{equation}

\noindent In Eq.~\eqref{eq:alphaconvFourier}, $\widehat{\omega}(\vect{k},a_H)$ is the Fourier transform of the non-singular distribution $\omega(\vect{x},a_H)$ with spreading radius $a_H$. With this, the corresponding long-range stress field $\tensor{\sigma}^{\rm long}(\vect{x},a_H)$ is obtained in the real-space by computing the inverse Fourier transform of Eq.~\eqref{eq:stressFourier}.

\subsection{Short-range stress contribution} \label{sec:shortrange}

Following the decomposition introduced in Eq.~\eqref{eq:stresstot}, a supplementary short-range contribution $\tensor{\sigma}^{\rm short}$ must be added to account for the stress component that is not captured by the grid-based approach alone. Taking advantage of the non-singular formulation, we note that, when considering an arbitrary dislocation segment $s$, the difference between the non-singular stress fields $\tensor{\sigma}^{s}(\vect{x},a_1)$ and $\tensor{\sigma}^{s}(\vect{x},a_2)$, obtained for different values of the core width $a_1$ and $a_2$, becomes vanishingly small at large distances $r$ away from the dislocation line. In practice, it is assumed that the difference becomes negligible at distances larger than a critical radius $r_c(a_1,a_2)$ chosen to satisfy a given error tolerance (see Appendix \ref{app:splittingerror}). With this, the short-range stress field correction $\tensor{\sigma}^{\rm short}(\vect{x},a_0,a_H)$ in Eq.~\eqref{eq:stresstot} is simply obtained as:

\begin{equation} \label{eq:stressshort}
\sigma_{ij}^{\rm short}(\vect{x},a_0,a_H) = \sum_n^{N_{\rm nei}} \left[ \sigma_{ij}^{n}(\vect{x},a_0) - \sigma_{ij}^{n}(\vect{x},a_H) \right]
\end{equation}

\noindent where the sum is carried over all neighbor segments $n$ whose minimum distance from field point $\vect{x}$ is closer than $r_c$, and $\tensor{\sigma}^{n}(\vect{x},a)$ is the non-singular stress field for which an analytical expression for straight segments in an isotropic media is given in \cite{Cai06}.

As detailed in Appendix \ref{app:splittingerror}, the error introduced in the splitting procedure presented in this work can be controlled by the choice of the critical radius $r_c$, which defines the set $\{n\} \in N_{\rm nei}$ of local neighbor segments to include in the short-range correction. For instance, for $a_0 = 1b$, it is observed that choosing $r_c = 4a_H$ corresponds to a introducing a maximum splitting error of approximately 5\% (see Appendix \ref{app:splittingerror}).

We note here that the treatment of the short-range contribution in this formulation offers two main advantages compared to the original DDD-FFT method. First, the nature of the splitting scheme eliminates the need for the ``capsule'' procedure (see \cite{Bertin15}) to identify portions of neighboring segments to be included in the short-range contribution, thereby simplifying the implementation of the method. This is because including more neighbors than necessary in the short-range contribution has no effect on the resulting stress field due to the convergence of the stress difference at distances beyond $r_c$. Second, we shall not be concerned about the translation tensor between both stress contributions in Eq.~\eqref{eq:stresstot} as previously discussed in \cite{Bertin15}, since the corrective term in Eq.~\eqref{eq:stressshort} only involves a stress difference in the current formulation.

\subsection{Numerical implementation}

Summarizing the preceding sections, the mechanical state and driving forces associated with dislocation lines present in the periodic simulation volume subjected to a mechanical load is numerically determined as follows:

\begin{enumerate}
\item Evaluate the dislocation density tensor $\tensor{\alpha}(\vect{x}^d)$ on a discrete grid $\Omega^d$ to obtain a continuous representation of the microstructure using the analytical approach presented in section \ref{sec:alphagrid}.
\item Fourier transform $\tensor{\alpha}(\vect{x}^d)$ and compute its non-singular spectral expression $\widehat{\tensor{\alpha}}^{\rm ns}(\vect{k}^d,a_H)$ and the corresponding $\widehat{\tensor{\chi}}(\vect{k}^d)$ tensor field using Eqs. \eqref{eq:chiFourier}-\eqref{eq:alphaconvFourier}.
\item Solve for the resulting stress field $\widehat{\tensor{\sigma}}^{\rm long}(\vect{k^d})$ using Eq. \eqref{eq:stressFourier} and compute its inverse Fourier transform to obtain the stress field $\tensor{\sigma}^{\rm long}(\vect{x^d})$ on the real-space grid $\Omega^d$.
\item Interpolate the stress field from the grid $\Omega^d$ to obtain the long-range contribution and add supplementary short-range contributions (see \S\ref{sec:stressdecomp}) to evaluate the total driving force acting along dislocation segments.
\end{enumerate}

Practically, the periodic simulation volume $\Omega$ is discretized into a uniform grid $\Omega^d$ of $N_d$ voxels, such that the Fourier transforms can be conveniently computed using the efficient FFT algorithm in $\mathcal{O}(N_d \log N_d)$ time. To further improve the efficiency of the method, a GPU implementation similar to that discussed in \cite{Bertin18a} has been developed for the spectral framework. In the context of DDD simulations, the calculation of the driving force on each dislocation segment is performed by integrating the total stress field acting along the segment line (see \S\ref{sec:ddd}). While the long-range contribution is numerically integrated by interpolation of the grid values, the short-range contribution is directly computed using existing analytical interactions forces expressions on pairs of neighbor segments \cite{Arsenlis07, Aubry14}.

Here, we also emphasize that the advantage of relying on the non-singular distribution expressed in Eq.~\eqref{eq:omega} is three-fold. First, as extensively discussed in this section, it provides a mathematical basis for splitting the stress contributions in a rigorous manner. Second, this splitting naturally results in a splitting of the force contributions, making the approach fully compatible with the efficient force-based subcycling time-integrator recently introduced in \cite{Sills16} (see \S\ref{sec:ddd}). Third, the non-singular kernel inherently provides a mechanism to spread the dislocation tensor field across voxels, thereby allowing to construct a field whose smoothness is fully controlled with parameter $a_H$. Of particular interest, since $a_H$ must be chosen larger than the grid spacing $H_i$ (see \S\ref{sec:longrange}), this approach naturally precludes the occurrence of spurious Gibbs oscillations that otherwise typically arise in the context of spectral methods.

\section{Validation of the model} \label{sec:validation}

\subsection{Dislocation stress field} \label{sec:stressfield}

To validate our model, we first consider the case of an infinite edge dislocation inserted into a periodic simulation box of length $L=5000b$, where $b=2.55 \times 10^{-10}$ m is the magnitude of the Burgers in Cu. The dislocation line is aligned along the $z$-direction and has a Burgers vector $\vect{b}=[b\,0\,0]$ along the $x$-direction. Elastic isotropy with shear modulus $\mu=56.4$ GPa and Poisson's ratio $\nu=0.324$ is used. A value of $a_0 = b$ for the physical dislocation core size is used, while the grid spreading parameter $a_H$ is set to twice the grid spacing in each direction.

\begin{figure}[t]
  \begin{minipage}[b]{0.5\linewidth}
  \begin{center}
    \includegraphics[scale=0.5]{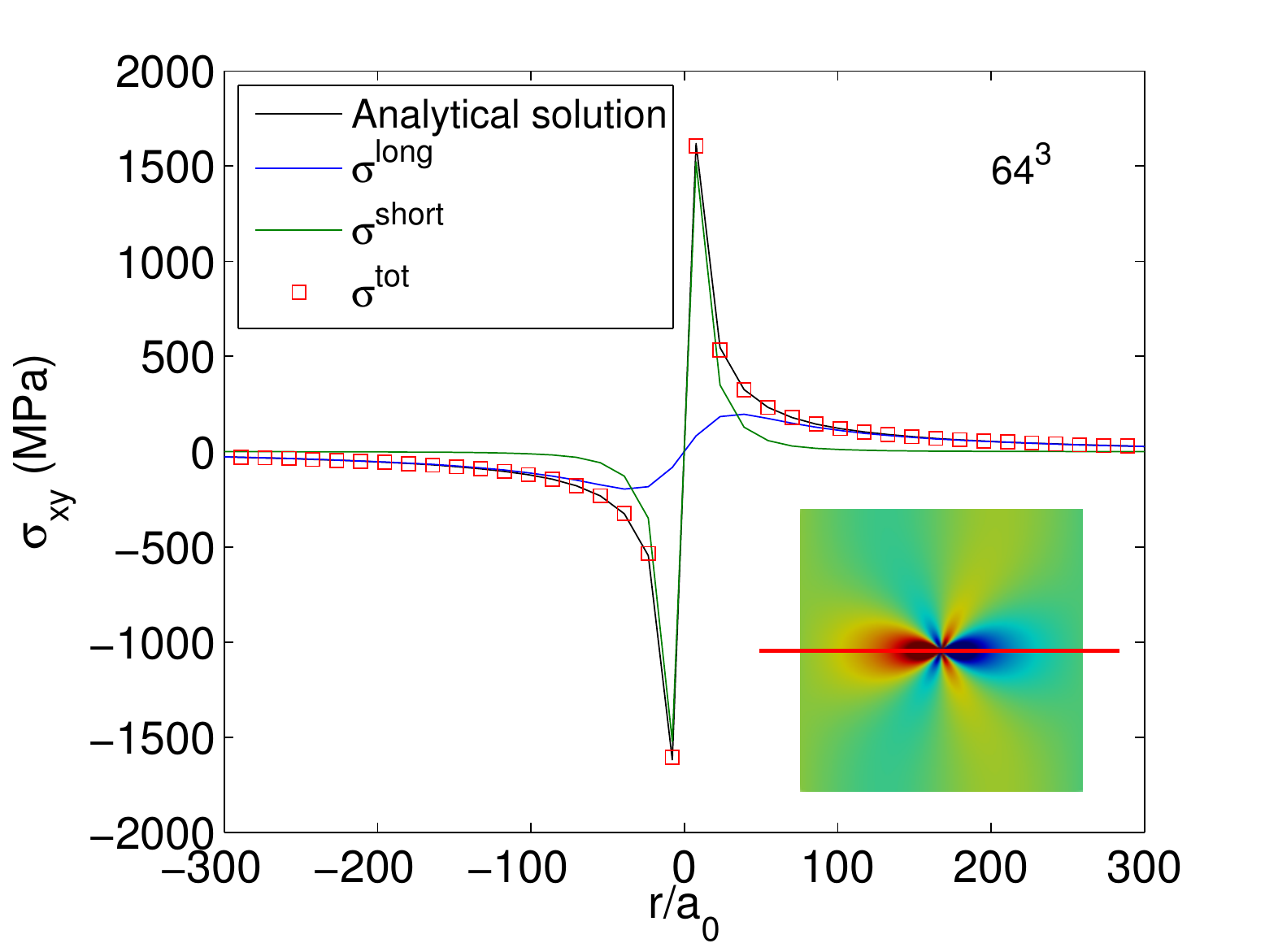} \\ (a)
  \end{center}
  \end{minipage}
  \begin{minipage}[b]{0.5\linewidth}
  \begin{center}
    \includegraphics[scale=0.5]{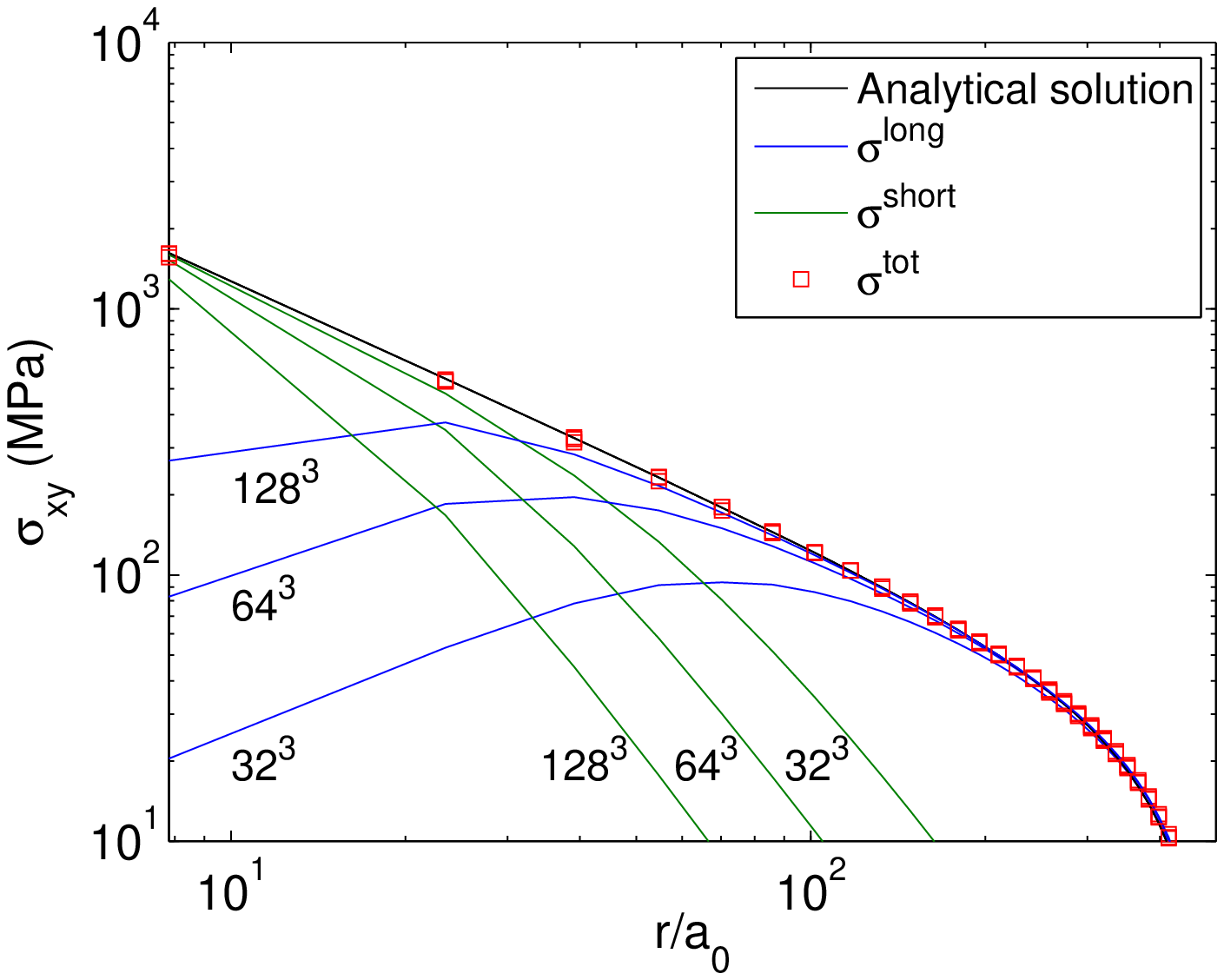} \\ (b)
  \end{center}
  \end{minipage}
  \caption{Stress field $\sigma_{xy}$ computed along a line passing through the center of an infinite edge dislocation line. (a) Decomposition of the long-range and short-range stress contributions for a spectral grid of resolution $N_d = 64^3$ voxels. (b) Stress contributions plotted on a log-log scale for different grid resolutions. The short-range contribution rapidly vanishes away from the dislocation line while the long-range component fades out in the core region. The total stress is plotted against the analytical non-singular solution to validate the accuracy of the splitting approach.}
  \label{fig:edgestressplitting}
\end{figure}

Fig.~\ref{fig:edgestressplitting}(a) shows the different $\sigma_{xy}$ stress contributions as calculated with the present method, along a $y$-line passing through the center of the dislocation, as shown in the inset. In this example, a grid resolution of $N_d=64^3$ voxels is used to calculate the long-range contribution $\sigma_{xy}^{\rm long}$, while the short-range contribution $\sigma_{xy}^{\rm short}$ is computed with Eq.~\eqref{eq:stressshort} using the non-singular stress expressions given in \cite{Cai06}. As expected, it is observed that, away from the dislocation line, the total stress field is mainly captured by the long-range contribution, while the short-range contribution becomes predominant in the core. When summing both contributions, it is shown that the total stress $\sigma_{xy}^{\rm tot}$ matches very well the analytical solution obtained from the non-singular expressions in \cite{Cai06} with core radius $a_0 = b$.

A more detailed examination of the splitting of the stress contributions is given in Fig. ~\ref{fig:edgestressplitting}(b) as a function of the grid resolution $N_d$. It is observed that, as the grid resolution is increased, the proportion of the long-range contribution to the total stress becomes more important overall, while the short-range contribution diminishes more rapidly away from the dislocation core. Clearly, this results from the fact that, as the grid resolution is increased, the number of neighbor segments to be included in the short-range contribution decreases. Consistent with this observation and Eq.~\eqref{eq:stressshort}, it has been verified that, when the grid is sufficiently fine such that $a_H = a_0$, the short-range contribution vanishes while the total stress is fully captured by the long-range spectral contribution. In addition, the results reported in Fig. ~\ref{fig:edgestressplitting}(b) demonstrate that, in all cases, the total stress agrees very well with the analytical solution, thereby demonstrating the independence of the stress field solution to the grid resolution.

\subsection{Frank-Read source activation} \label{sec:frsactivation}

As another benchmark, we now investigate the critical stress $\tau_{\rm act}$ required to activate a Frank-Read source. The Frank-Read source activation is of particular importance since it is one of the unit-processes that govern the plastic flow in single crystals.

\begin{figure}[t]
  \begin{minipage}[b]{0.5\linewidth}
  \begin{center}
    \includegraphics[scale=0.55]{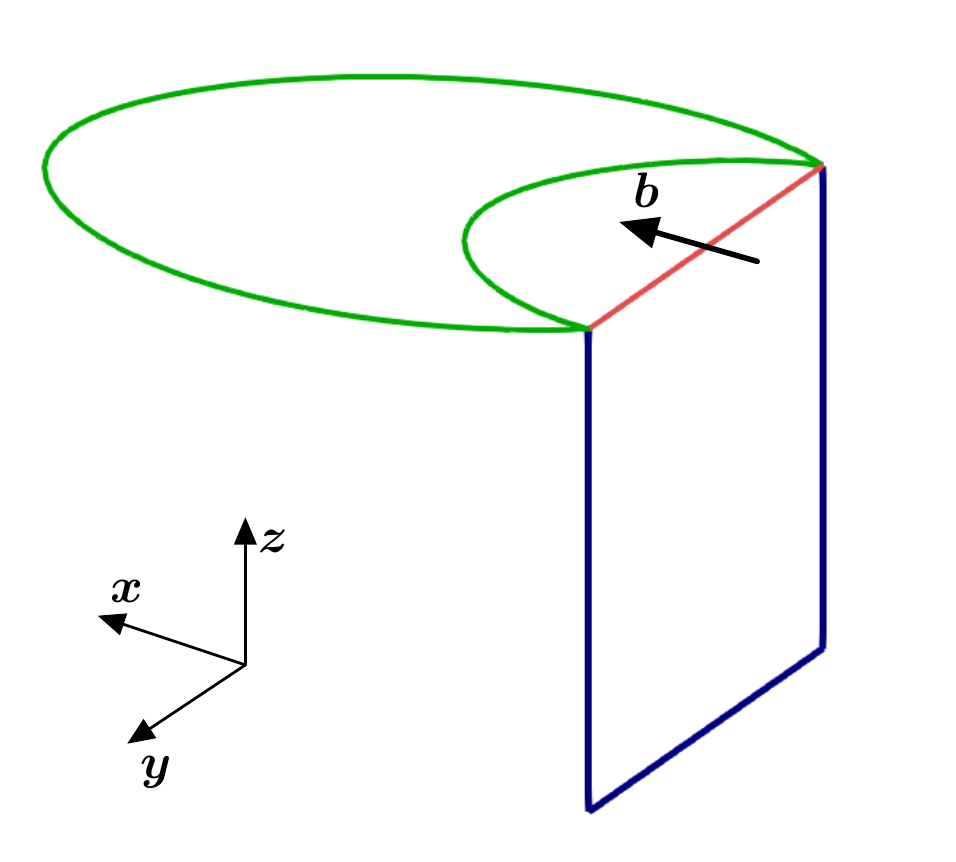} \\ (a)
  \end{center}
  \end{minipage}
  \begin{minipage}[b]{0.5\linewidth}
  \begin{center}
    \includegraphics[scale=0.4]{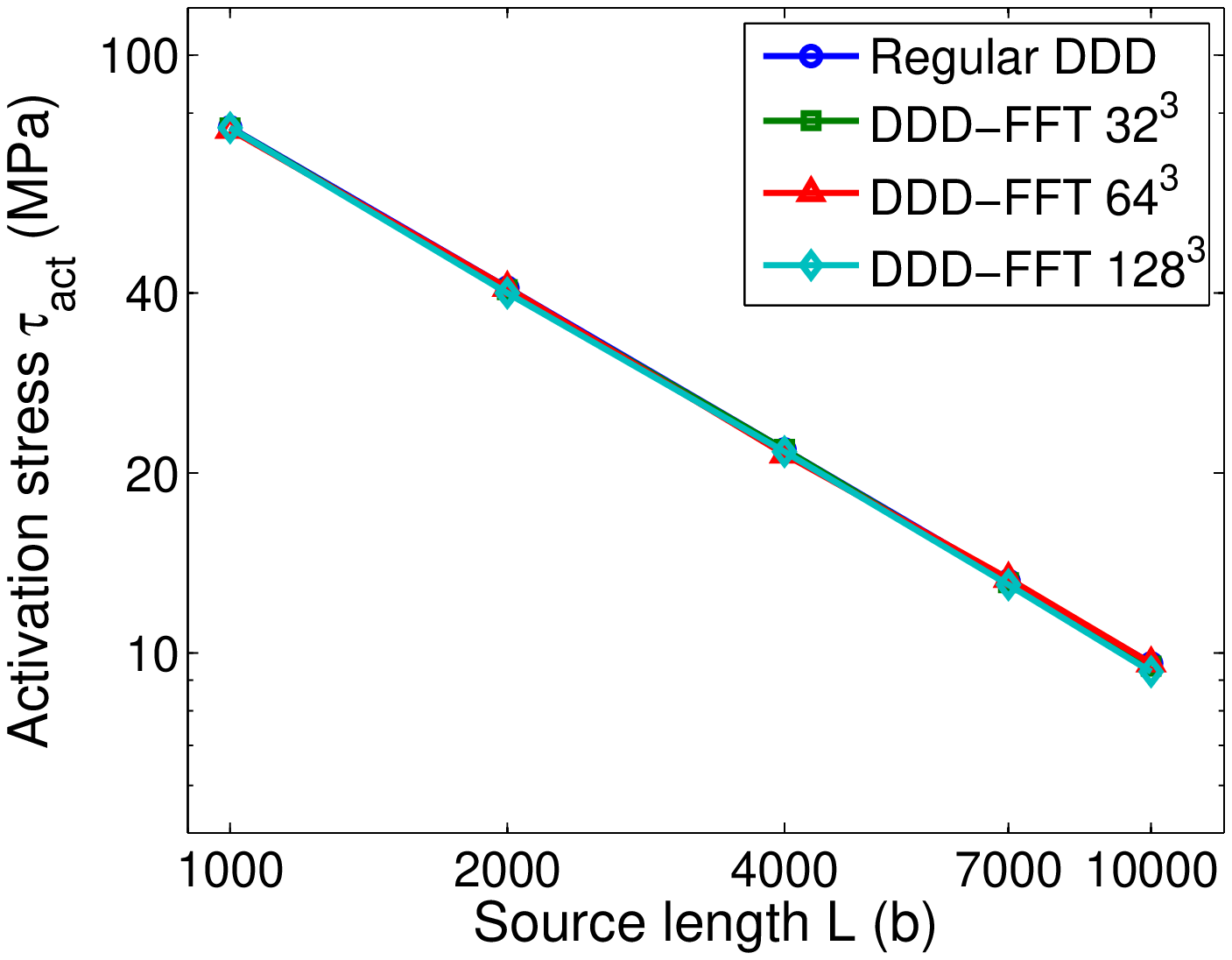} \\ (b)
  \end{center}
  \end{minipage}
  \caption{(a) Setting used to determine the activation stress $\tau_{\rm act}$ of a Frank-Read source. The source is created by inserting a prismatic dislocation loop composed of four arms of length $L$ with Burgers vector $\vect{b} = [b\,0\,0]$ along the $x$-direction. Three of its arms (blue) are immobilized; the remaining mobile arm (red) is pinned at both ends. When subjected to an applied resolved shear stress $\tau_{xz}$, the mobile arm acts as an edge source that glides on the $[0\,0\,1]$ plane, as shown in green. (b) Activation stress $\tau_{\rm act}$ obtained when using the current DDD-FFT approach for different resolutions compared to that obtained when using the original DDD approach.}
  \label{fig:frsactivation}
\end{figure}

We employ the setting depicted in Fig.~\ref{fig:frsactivation}(a). To avoid spurious effects associated with evaluating stress fields on terminating segments, the Frank-Read source is created by considering a prismatic dislocation loop composed of four arms of length $L$ with Burgers vector $\vect{b} = [b\,0\,0]$ along the $x$-direction. The three lower arms (blue) of the loop are immobilized such that the remaining mobile arm (red) acts as an edge source of length $L$ pinned at both ends that can glide on the $[0\,0\,1]$ plane, as shown in Fig.~\ref{fig:frsactivation}(a). The activation stress $\tau_{\rm act}$ is determined by increasing the applied resolved shear stress $\tau_{xz}$ by increments of $0.1$ MPa until activation of the source is observed. The material parameters for Cu given in section \ref{sec:stressfield} are used.

Activation stresses $\tau_{\rm act}$ obtained for different source lengths $L$ ranging from $1000\,b$ to $10000\,b$ are reported in Fig.~\ref{fig:frsactivation}(b) for different grid resolutions $N_d = 32^3$ to $128^3$ voxels to evaluate long-range stress contributions. It is shown that the results obtained are almost insensitive to the choice of the grid resolution $N_d$, thereby demonstrating the consistency of the DDD-FFT approach and the validity of our submesh resolution procedure. For reference, the results are also compared with activation stresses obtained when using an original DDD approach, in which nodal forces are obtained by explicitely summing up the contribution of every dislocation segment using non-singular analytical expressions (e.g. as in \cite{Arsenlis07}). A very good agreement is observed between both methods, thereby validating our approach in dynamics cases.

\section{Applications} \label{sec:applications}

\subsection{Discrete dislocation dynamics} \label{sec:ddd}

Except in section \ref{sec:frsactivation}, the focus up to this point was primary placed on the calculation of the stress field resulting from a distribution of dislocation lines. In the context of DDD simulations, the calculation of the total stress is a critical step in that it directly allows to determine the driving force acting along dislocation segments, which is required to evaluate dislocations velocities. Specifically, in the nodal approach, where the dislocation network is decomposed into a series of $N_{\rm seg}$ inter-connected dislocation segments (e.g. see \cite{Zbib98, Arsenlis07}), the Peach-Koelher force,

\begin{equation}
\vect{f}^{\rm pk} = \left( \tensor{\sigma}^{\rm tot} \cdot \vect{b} \right) \times \vect{t} \, ,
\end{equation}

\noindent acting along dislocation line with Burgers vector $\vect{b}$ and line tangent $\vect{t}$, is integrated along the segment such as to obtain the total force acting on each node of the discretized network. Using the splitting introduced in Eq.~\eqref{eq:stresstot}, the force $\vect{f}^s_i$ on straight segment $s = ij$ (i.e. defined between end nodes $i$ and $j$) acting at node $i$ is obtained as:

\begin{align} \label{eq:forcetot}
\vect{f}^s_i &= \int_{L^s} N_i(\vect{x})  \left[ \left( \tensor{\sigma}^{\rm tot}(\vect{x},a_0) \cdot \vect{b} \right) \times \vect{t} \right] dL(\vect{x}) \nonumber\\
&= \int_{L^s} N_i(\vect{x})  \left[ \left( \tensor{\sigma}^{\rm long}(\vect{x},a_H) \cdot \vect{b} \right) \times \vect{t} \right] dL(\vect{x}) \nonumber\\
&+ \int_{L^s} N_i(\vect{x})  \left[ \left( \tensor{\sigma}^{\rm short}(\vect{x},a_0,a_H) \cdot \vect{b} \right) \times \vect{t} \right] dL(\vect{x}) \nonumber\\
&= \vect{f}^{s,{\rm long}}_i(a_H) + \vect{f}^{s,{\rm short}}_i(a_0,a_H)
\end{align}

\noindent where $N_i$ is the shape function associated with node $i$, and $L^s$ denotes the length of segment $s$. While the long-range force contribution is obtained by interpolation of the long-range stress grid values, the short-range component is calculated from explicit expressions derived for pairs of segments. Specifically, following Eq.~\eqref{eq:stressshort}, the short-range force is calculated as:

\begin{equation} \label{eq:forceshort}
\vect{f}^{s,{\rm short}}_i(a_0,a_H) = \sum_{kl}^{N_{\rm nei}} \left[ \vect{f}^{s,kl}_{i}(a_0) - \vect{f}^{s,kl}_{i}(a_H) \right]
\end{equation}

\noindent where the sum is carried over all neighboring segments $kl$ whose minimum distance from segment $s$ is smaller than the critical splitting radius $r_c(a_0,a_H)$ defined in \S\ref{sec:shortrange}. In Eq.~\eqref{eq:forceshort}, $\vect{f}^{s,kl}_{i}(a)$ denotes the interaction force between segments $s=ij$ and $kl$ acting at node $i$ with core radius $a$, for which an analytical non-singular expression for isotropic elasticity is given in \cite{Arsenlis07}, and a semi-analytical formulation for anisotropic elasticity is developed in \cite{Aubry13}.

Interestingly, we note here that the smooth force decomposition in Eq.~\eqref{eq:forcetot} additionally offers another significant benefit in the context of the simulation of dynamical systems. As extensively discussed in \cite{Sills14,Sills16}, the time-integration procedure in DDD is difficult due to the intrinsically stiff nature of the problem. As such, conventional time-integrators (e.g. Heun or Runge-Kutta-Fehlberg schemes) are required to take increasingly small time step sizes as the dislocation network is being refined with strain, thereby severely limiting the amount of strain that can be reached in practice. To circumvent this issue, a new subcycling integrator was recently developed, allowing to speed-up the computation by up to two orders of magnitude \cite{Sills16}. In order to achieve so, the core idea of the approach relies in separating the different interaction forces in several groups based on their stiffness. From there, each group is integrated separately with a proper time step size in a sequential fashion. The force splitting scheme introduced in this work precisely allows for such a separation of force interactions, thereby making the current approach naturally compatible with the force-based subcycling approach introduced in \cite{Sills16}. 

\begin{figure}[t]
  \begin{minipage}[b]{0.5\linewidth}
  \begin{center}
    \includegraphics[scale=0.4]{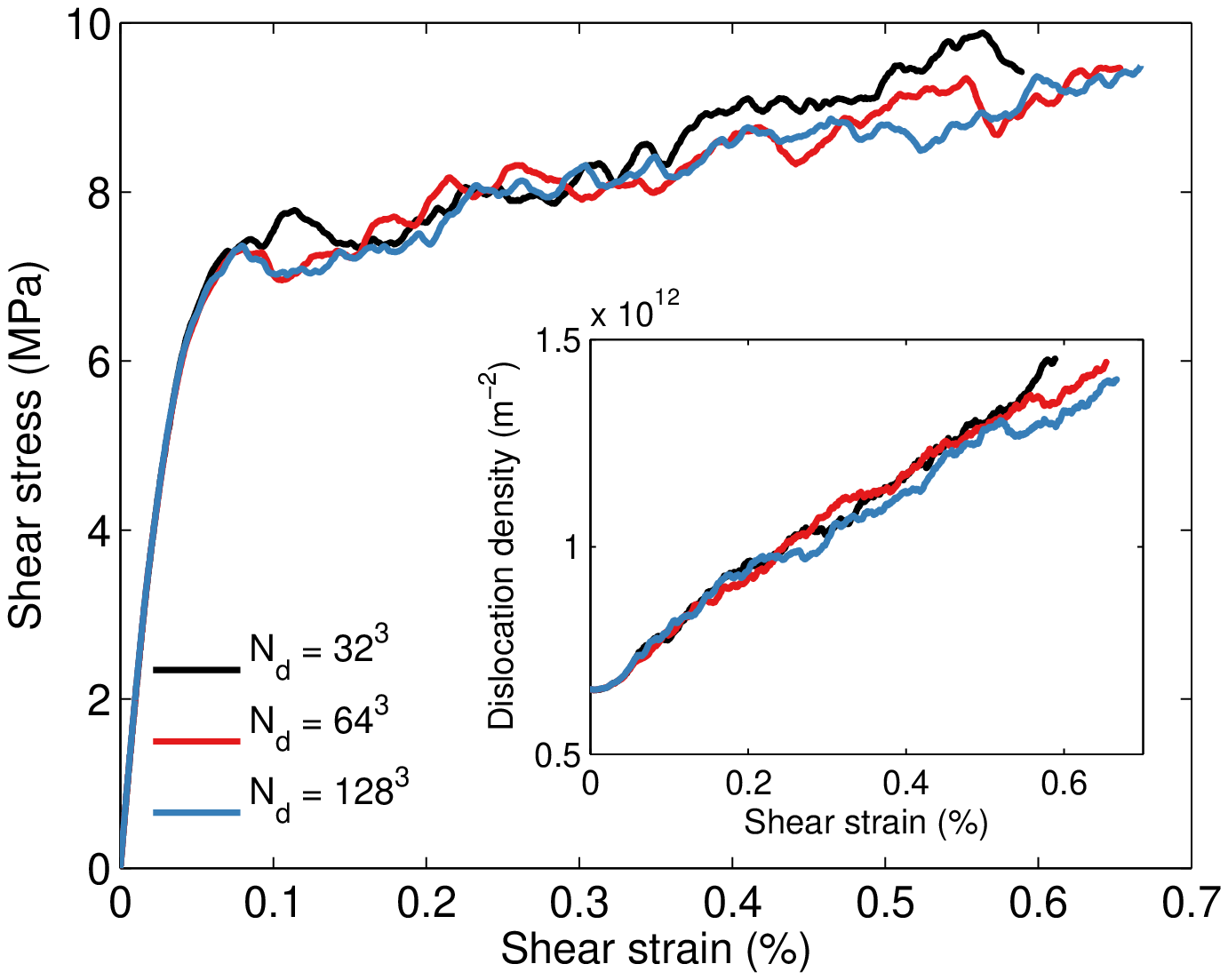} \\ (a)
  \end{center}
  \end{minipage}
  \begin{minipage}[b]{0.5\linewidth}
  \begin{center}
    \includegraphics[scale=0.25]{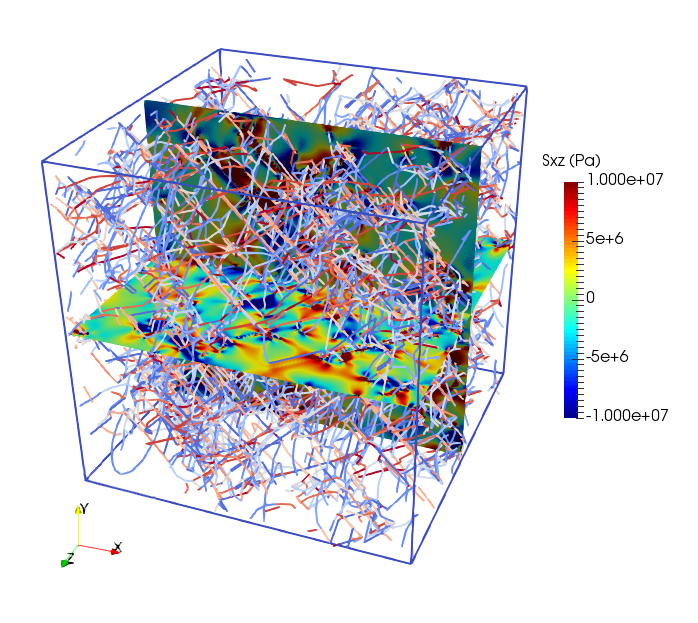} \\ \vspace*{-0.3cm} (b)
  \end{center}
  \end{minipage}
  \caption{(a) Example of shear stress-strain curves predicted by the DDD-FFT model for different spectral grid resolutions $N_d$ for a $(15 \, {\rm \mu m})^3$ FCC Cu periodic structure. (b) Example of a dislocation microstructure at $\gamma = 0.5\%$ shear strain and the corresponding $\sigma_{xz}$ total stress component plotted on two different planes.}
  \label{fig:dddresults}
\end{figure}

In this work, our new DDD-FFT framework was coupled with the force-based subcycling approach. Following the implementation proposed in \cite{Sills16}, we find that best performance is achieved by firstly time-integrating the long-range contribution alone so as to determine the global time step size, while further separating the short-range contribution into four supplementary groups based on their distance (see \cite{Sills16} for further details). An example of the stress-strain curves obtained from a work-hardening simulation in FCC Cu using the current approach is presented in Fig.~\ref{fig:dddresults}. In this simulation, an initially relaxed $(15\mu m)^3$ microstructure with dislocation density $\rho_0 = 7 \times 10^{11}$ m$^{-2}$ was loaded along the $[0\,0\,1]$ direction at a strain rate of $\dot{\epsilon} = 10^3$ s$^{-1}$. The long-range component was calculated using a grid resolutions ranging from $N_d = 32^3$ to $128^3$ voxels. As expected, it is observed that the predicted response is very consistent between all resolutions, thereby illustrating the robustness of our approach.

Note here that the force decomposition scheme also allows to efficiently recompute forces on nodes for which the local topology has changed, without the need to recompute the long-range component. For instance, this allows to calculate the dissipation associated with different trials configurations during the node splitting procedure  (see \cite{Arsenlis07}) in an efficient manner.

\subsection{Connection with Field Dislocation Mechanics}

At his core, the FDM approach proposes a full-field model for small-scale plasticity in which the dislocation content is represented using the dislocation density tensor. Building upon conventional plasticity frameworks, the model aims at evolving the dislocation tensor, essentially by relying on (i) the conservation of the Burgers vector, and (ii) thermodynamically-based considerations that require the dissipation, induced by dislocation shearing the crystal, to remain positive at all times.

Theoretically, such a model contains all the ingredients required to provide a self-consistent, closed theory. In practice however, one of the principal limitations  of the FDM currently lies in the difficulty associated with the determination of the velocity field required to evolve the density tensor field with deformation. Specifically, the building of a general connection between the internal stress fields (driving force) and the resulting velocity operator appears as a daunting task, even under simplifying assumptions. As a result, the FDM, in its original form where individual dislocations are resolved, was solely applied to the case of one-dimensional evolution problems \cite{XZhang15}. In higher-dimensions, dynamical approaches typically assume a particular form for the velocity functional, or simply resort to imposing constant velocity fields \cite{Djaka15}.

Concurrently to the development of the FDM, a coarse-grained version of the approach, called Phenomenological Macroscopic FDM (PMFDM), was introduced \cite{Acharya06}. While the original intent of this approach was to propose a model in which the FDM constitutive framework could be used to model spatial and temporal averaged fields evolution, it concurrently allows to circumvent the difficulty associated with the treatment of the velocity; in the PMFDM, the dislocation velocity field is related in a constitutive fashion to the shear strain rate, which itself is specified phenomenologically, e.g. chosen in the form of a power-law \cite{Acharya06}, as in a vast majority of crystal-plasticity models. With this however, the velocity becomes intrinsically disconnected from the internal microstructure, and the original goal of the approach to capture internal stresses associated with the presence of geometrically necessary dislocations becomes ambiguous. More importantly, the ability of such a method to generate dislocation density fields that capture the complexity of dislocation networks observed in DDD and atomistic simulations remains questionable, specifically because the evolution of the system results from a dynamical process in which short-range and core interactions are no longer explicitly resolved.

In this context, the method developed in this work can be regarded as a complementary approach to both the FDM and PMFDM models. Essentially, it provides an approach in which the evolution of the dislocation density tensor is directly provided by the motion of dislocation lines explicitly computed with a DDD model. In addition, the non-singular stress decomposition scheme provides a solution to complement current FDM implementations (e.g. \cite{Brenner14, Berbenni14}) with submesh resolution. Of particular interest, the current DDD-FFT method therefore offers an opportunity to compare, assess, and inform evolution laws used in current FDM-based approaches.

\begin{figure}[t]
  \begin{minipage}[b]{0.24\linewidth}
  \begin{center}
    \includegraphics[scale=0.25]{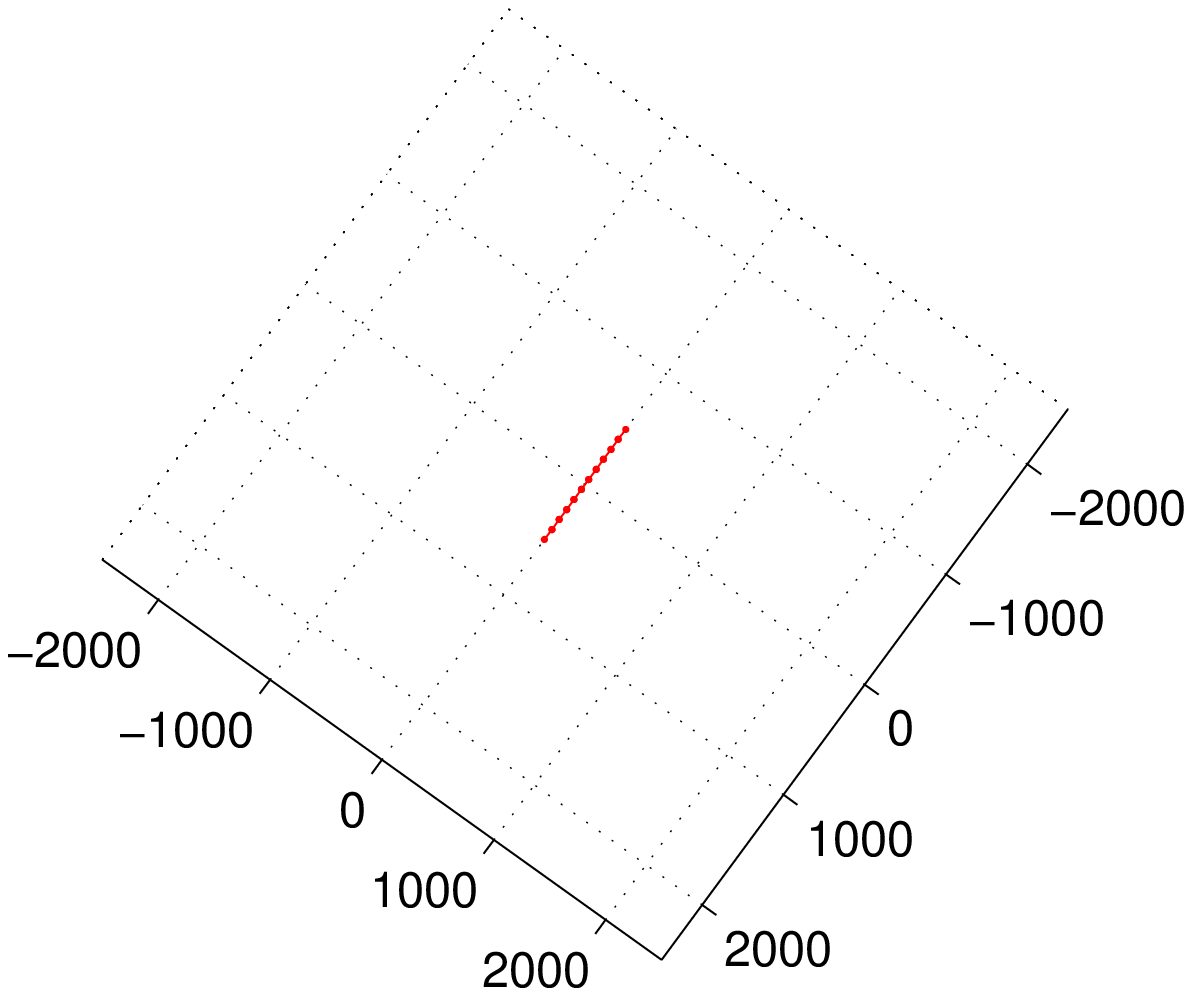}
  \end{center}
  \end{minipage}
  \begin{minipage}[b]{0.24\linewidth}
  \begin{center}
    \includegraphics[scale=0.25]{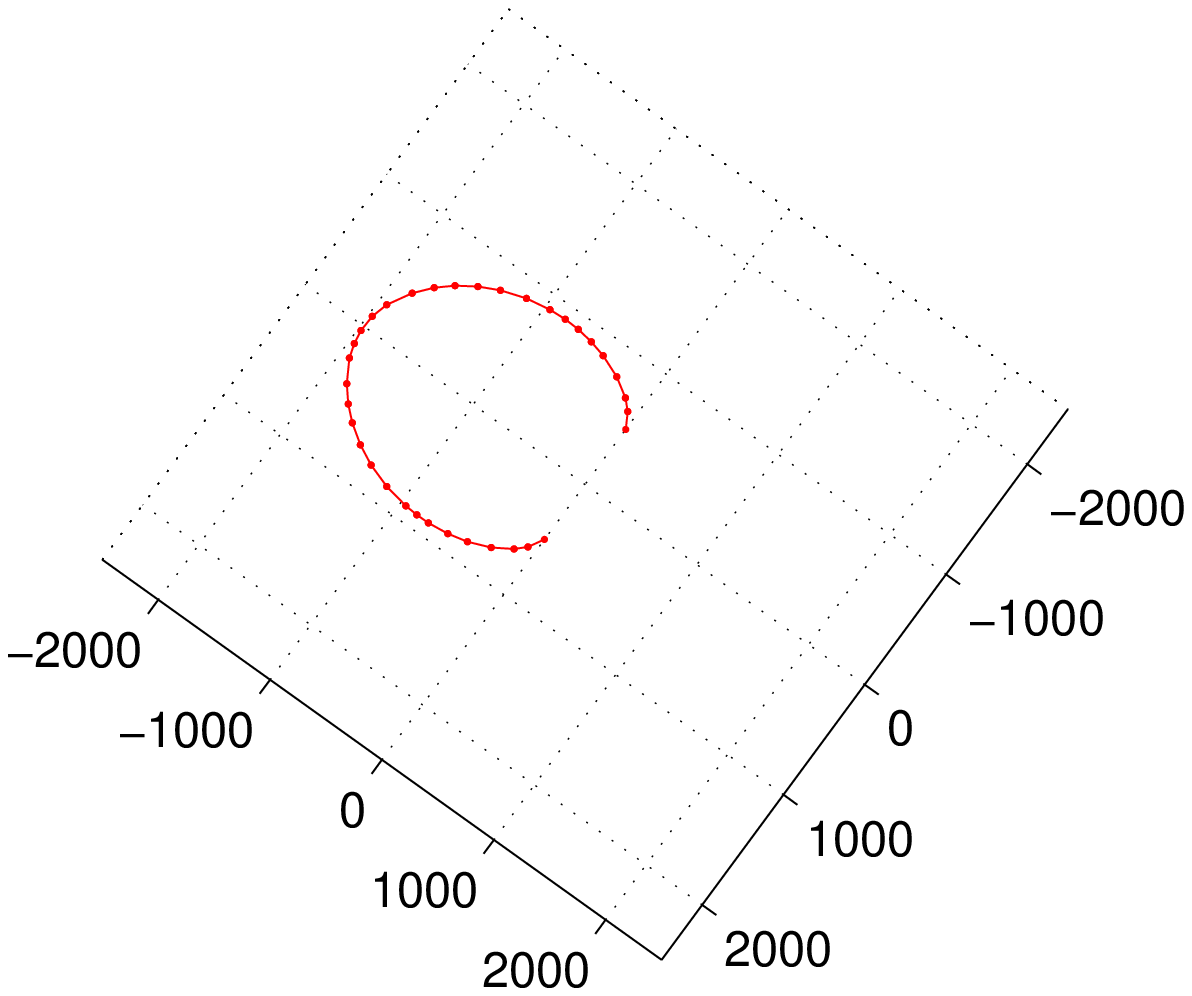}
  \end{center}
  \end{minipage}
  \begin{minipage}[b]{0.24\linewidth}
  \begin{center}
    \includegraphics[scale=0.25]{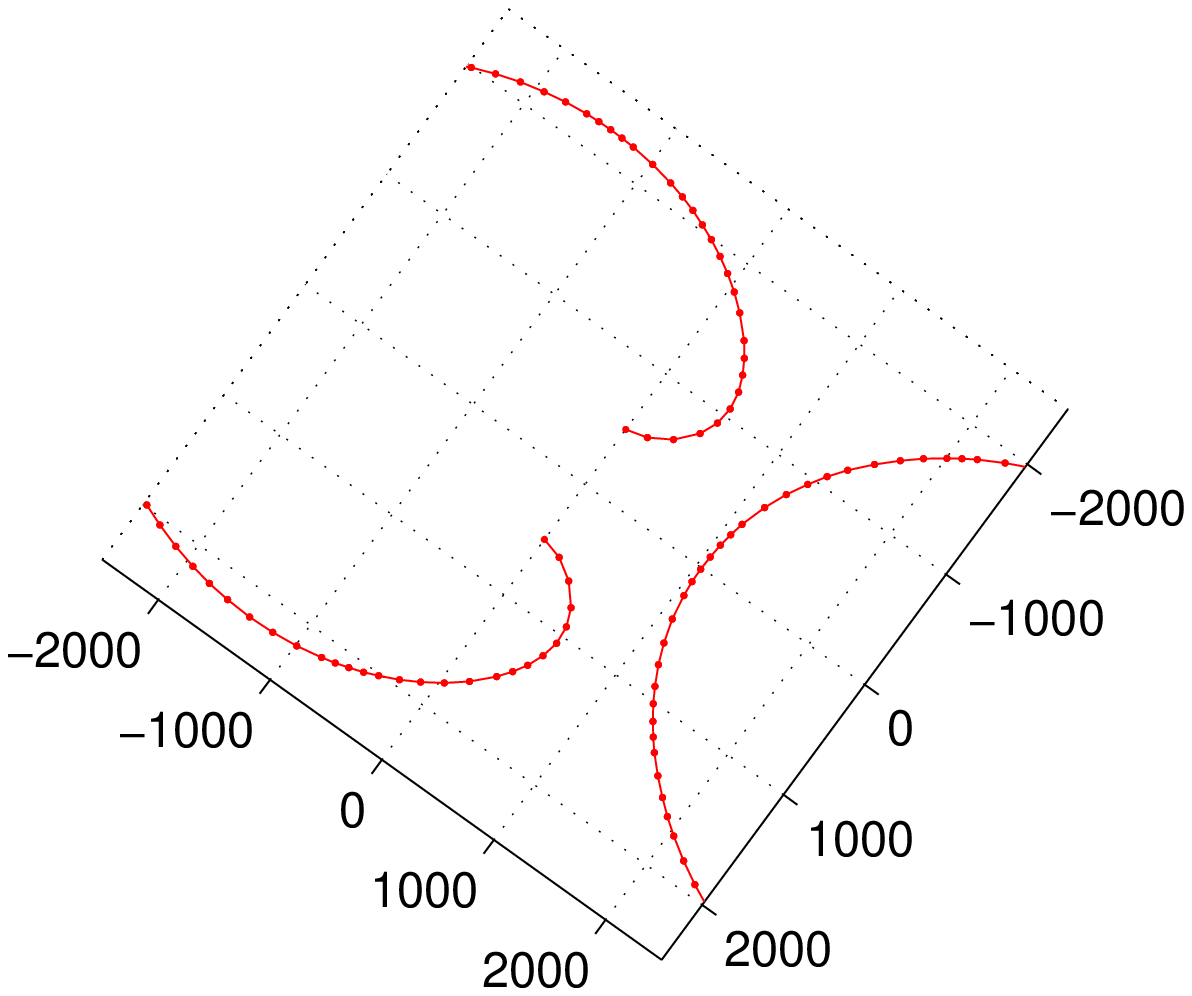}
  \end{center}
  \end{minipage}
  \begin{minipage}[b]{0.24\linewidth}
  \begin{center}
    \includegraphics[scale=0.25]{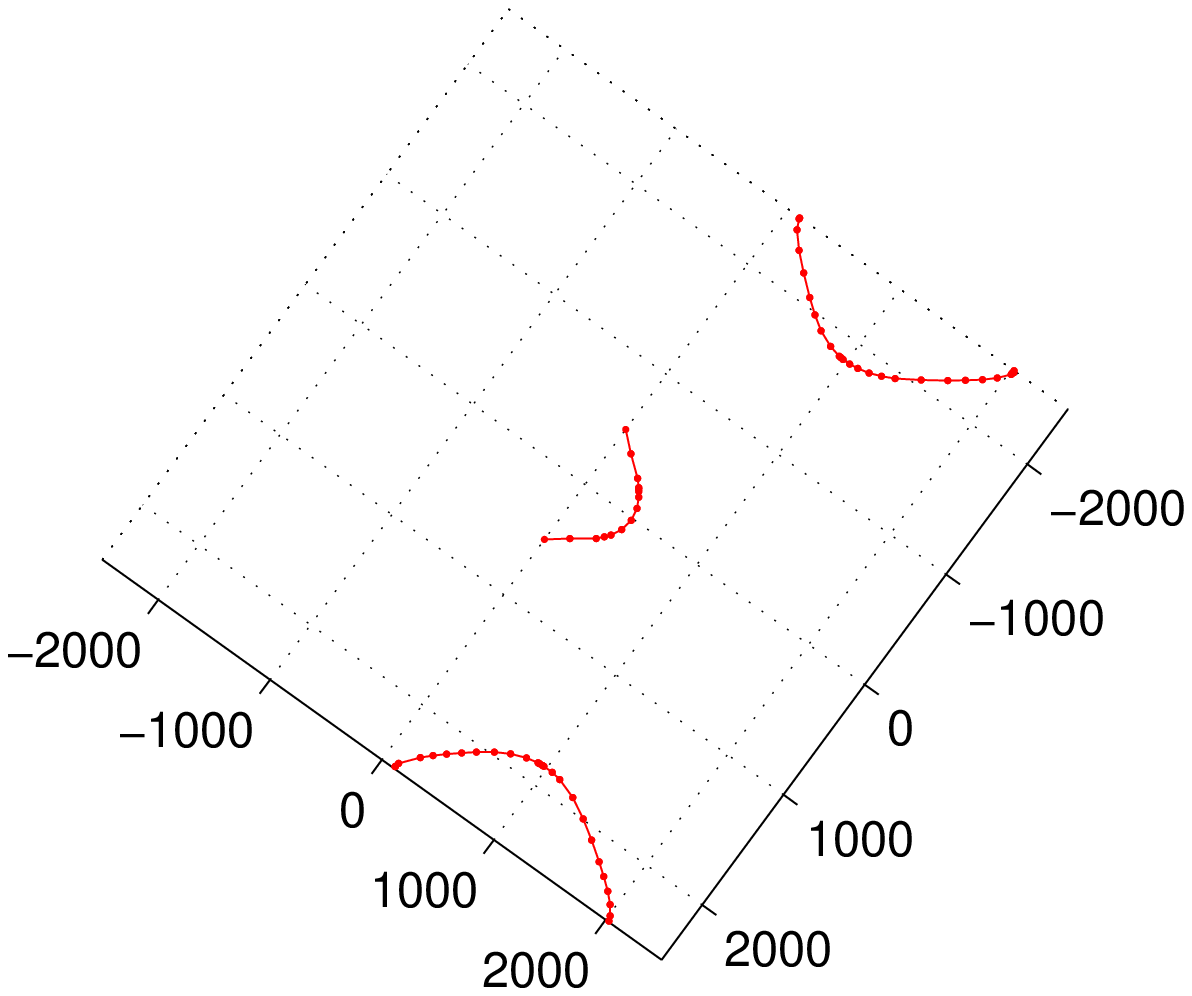}
  \end{center}
  \end{minipage}
  \begin{minipage}[b]{0.24\linewidth}
  \begin{center}
    \includegraphics[scale=0.25]{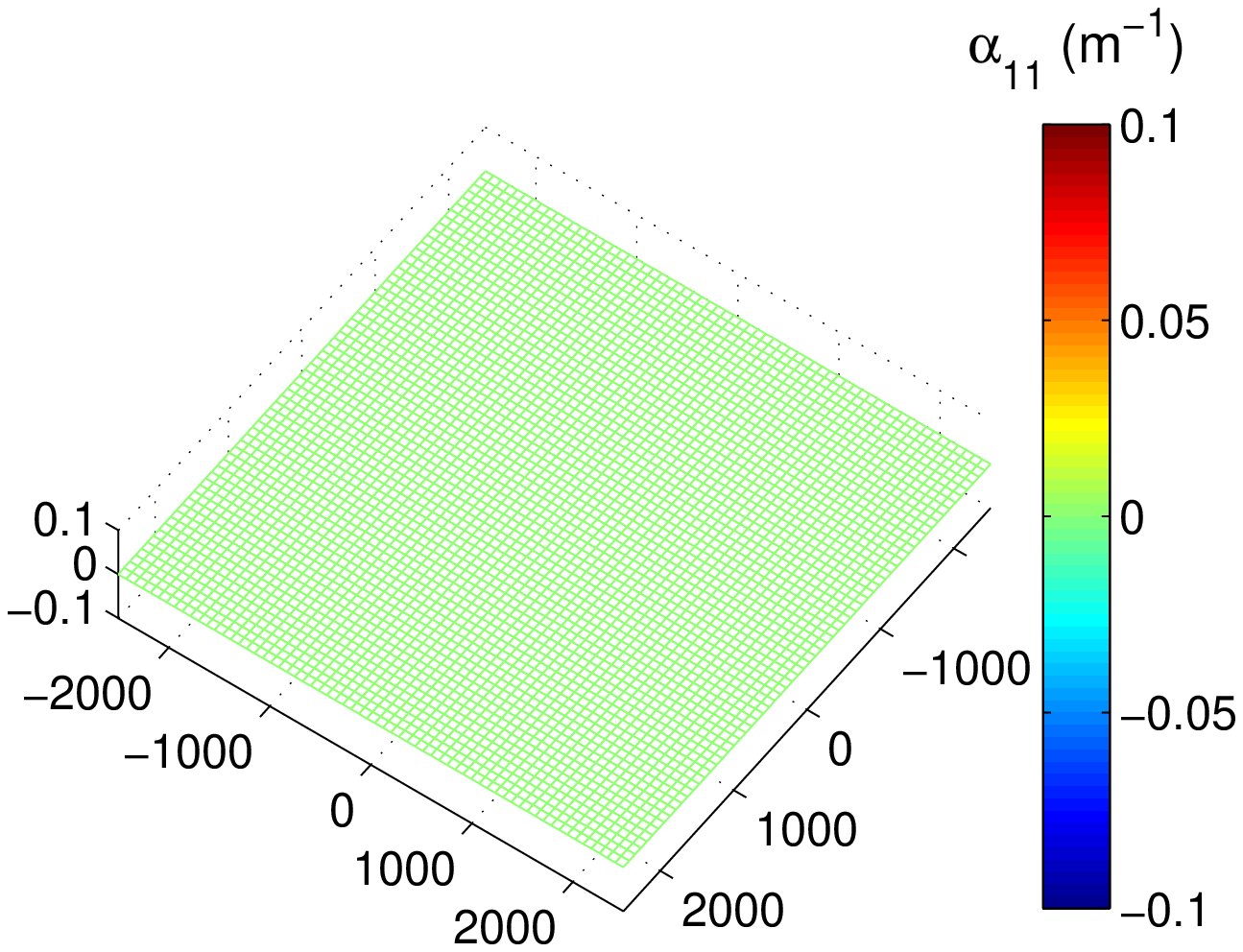}
  \end{center}
  \end{minipage}
  \begin{minipage}[b]{0.24\linewidth}
  \begin{center}
    \includegraphics[scale=0.25]{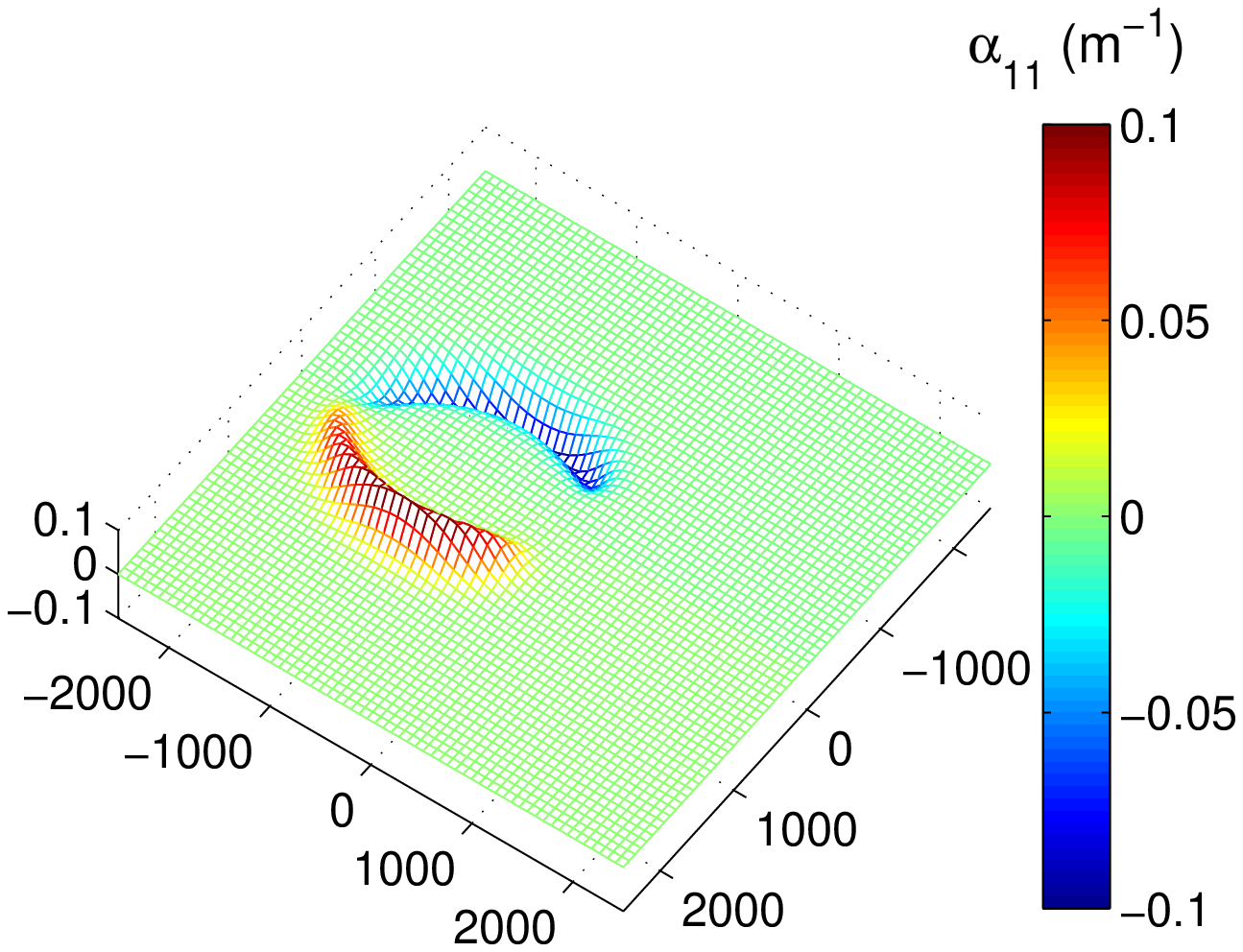}
  \end{center}
  \end{minipage}
  \begin{minipage}[b]{0.24\linewidth}
  \begin{center}
    \includegraphics[scale=0.25]{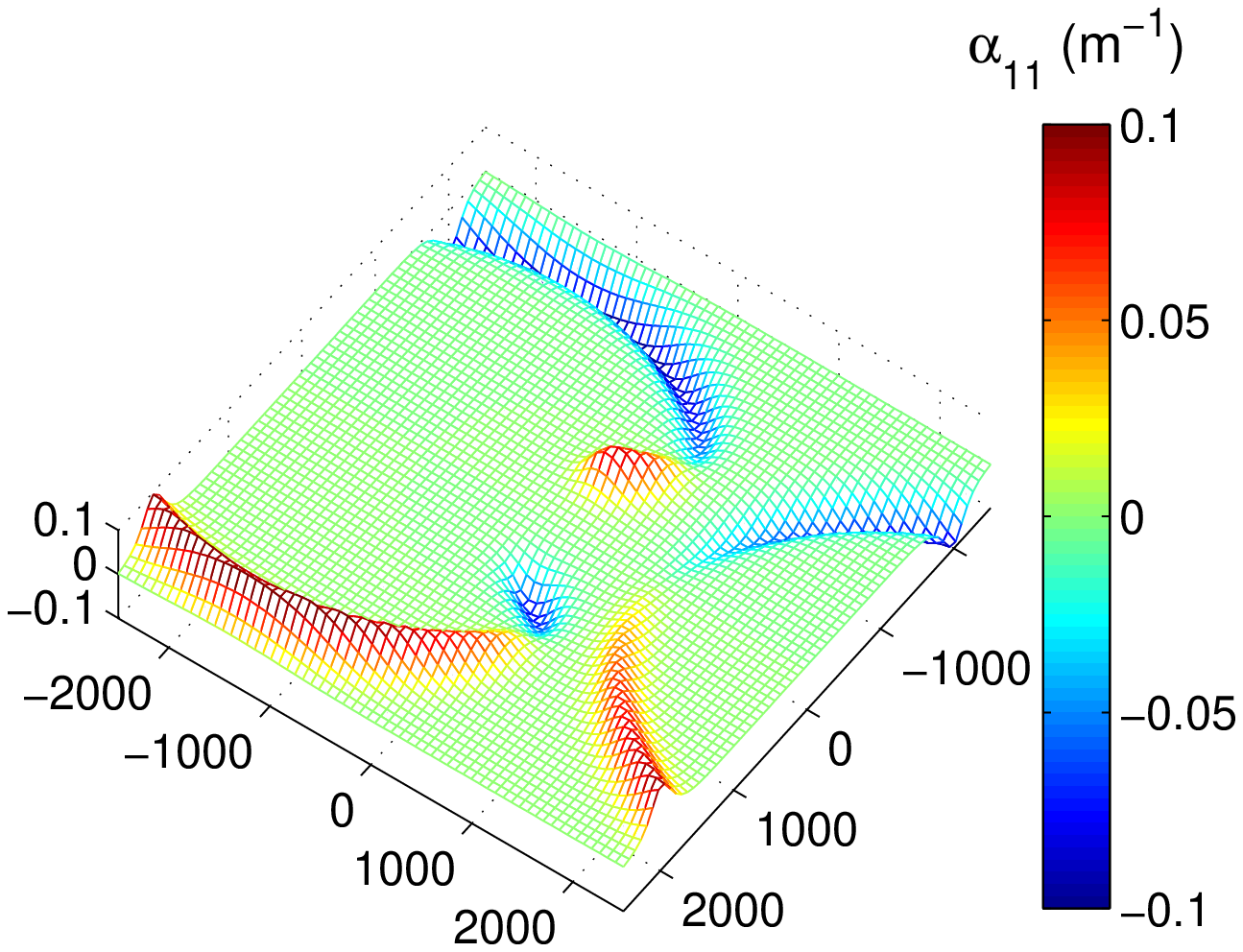}
  \end{center}
  \end{minipage}
  \begin{minipage}[b]{0.24\linewidth}
  \begin{center}
    \includegraphics[scale=0.25]{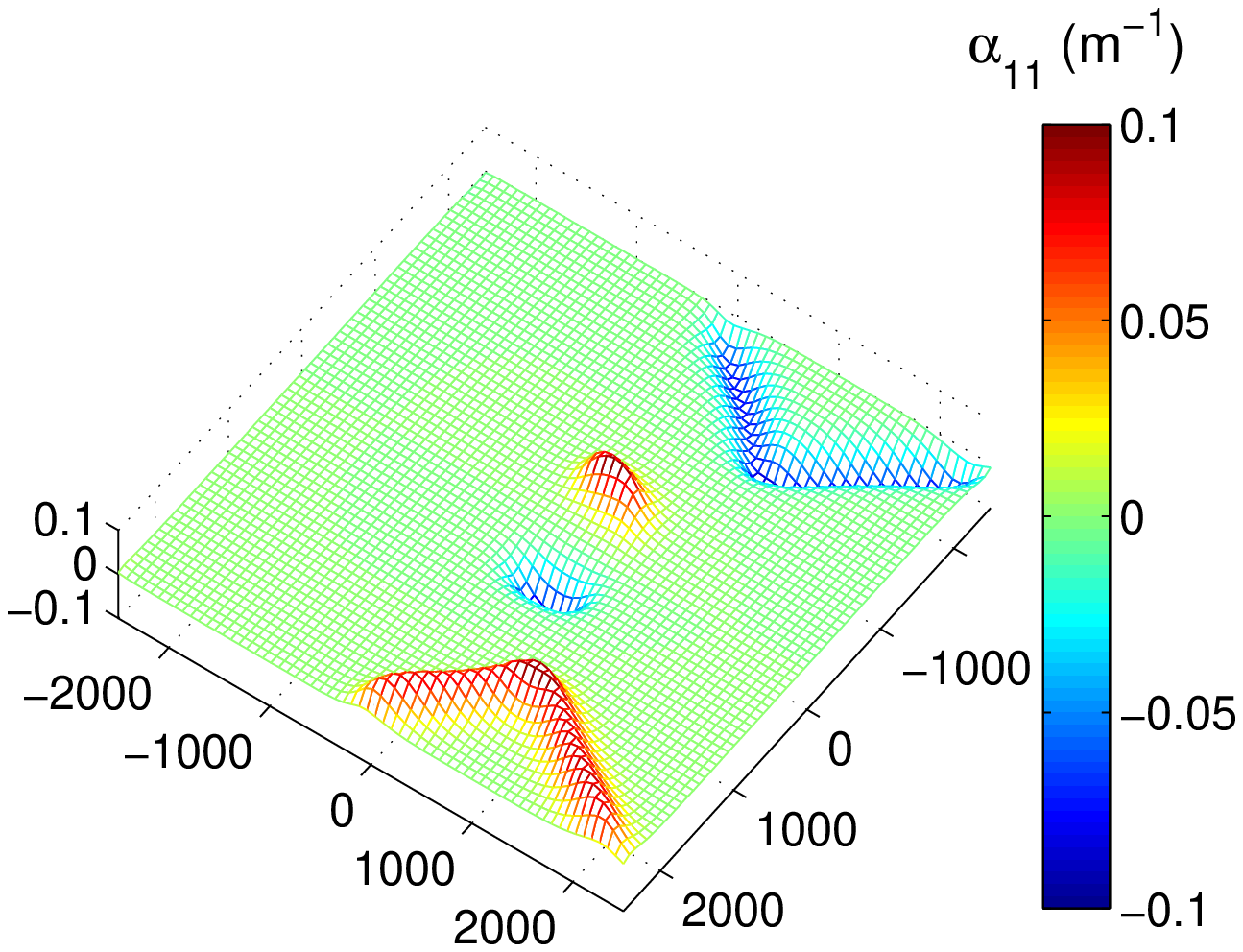}
  \end{center}
  \end{minipage}
  \begin{minipage}[b]{0.24\linewidth}
  \begin{center}
    \includegraphics[scale=0.25]{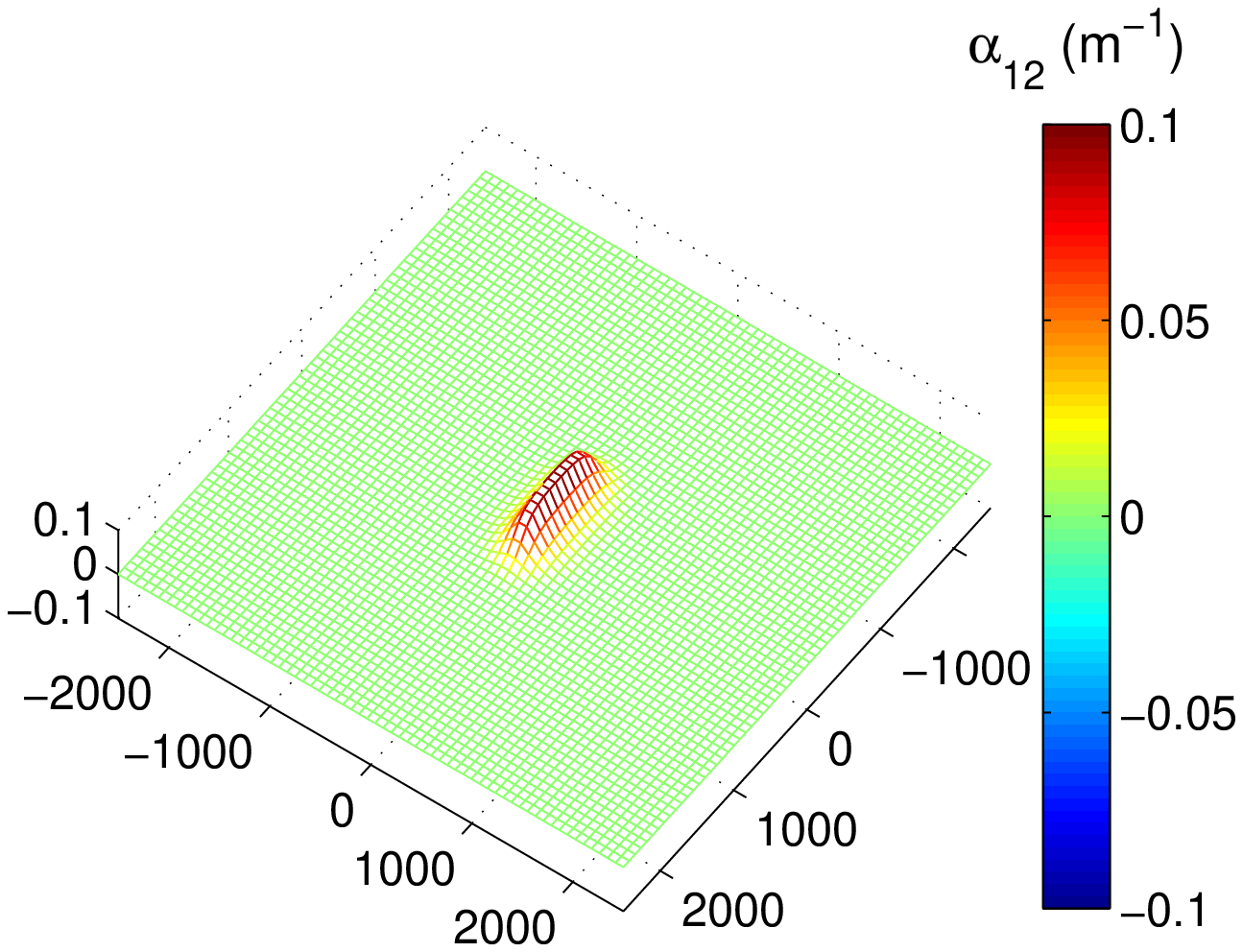} \\ (a) Step 1
  \end{center}
  \end{minipage}
  \begin{minipage}[b]{0.24\linewidth}
  \begin{center}
    \includegraphics[scale=0.25]{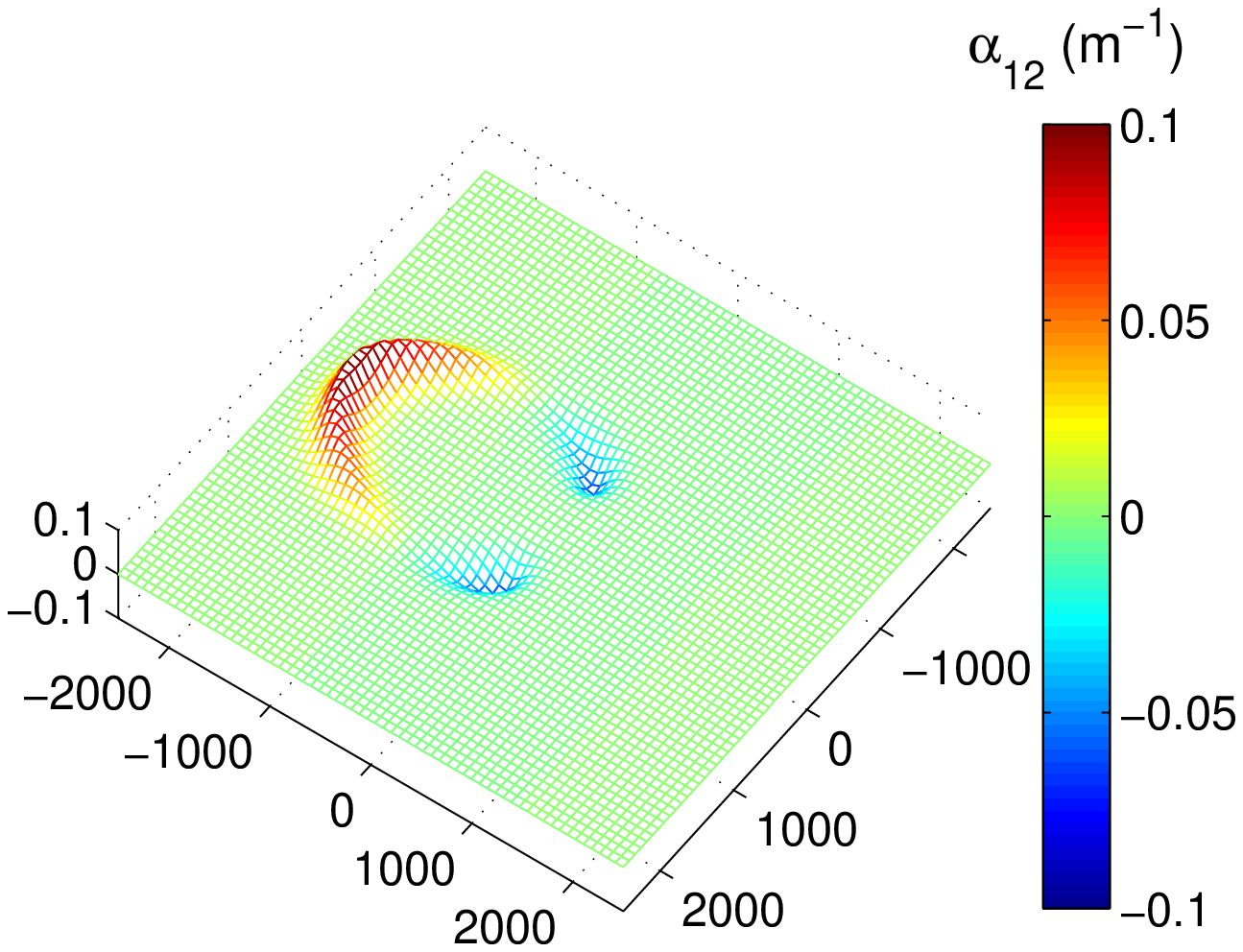} \\ (b) Step 100
  \end{center}
  \end{minipage}
  \begin{minipage}[b]{0.24\linewidth}
  \begin{center}
    \includegraphics[scale=0.25]{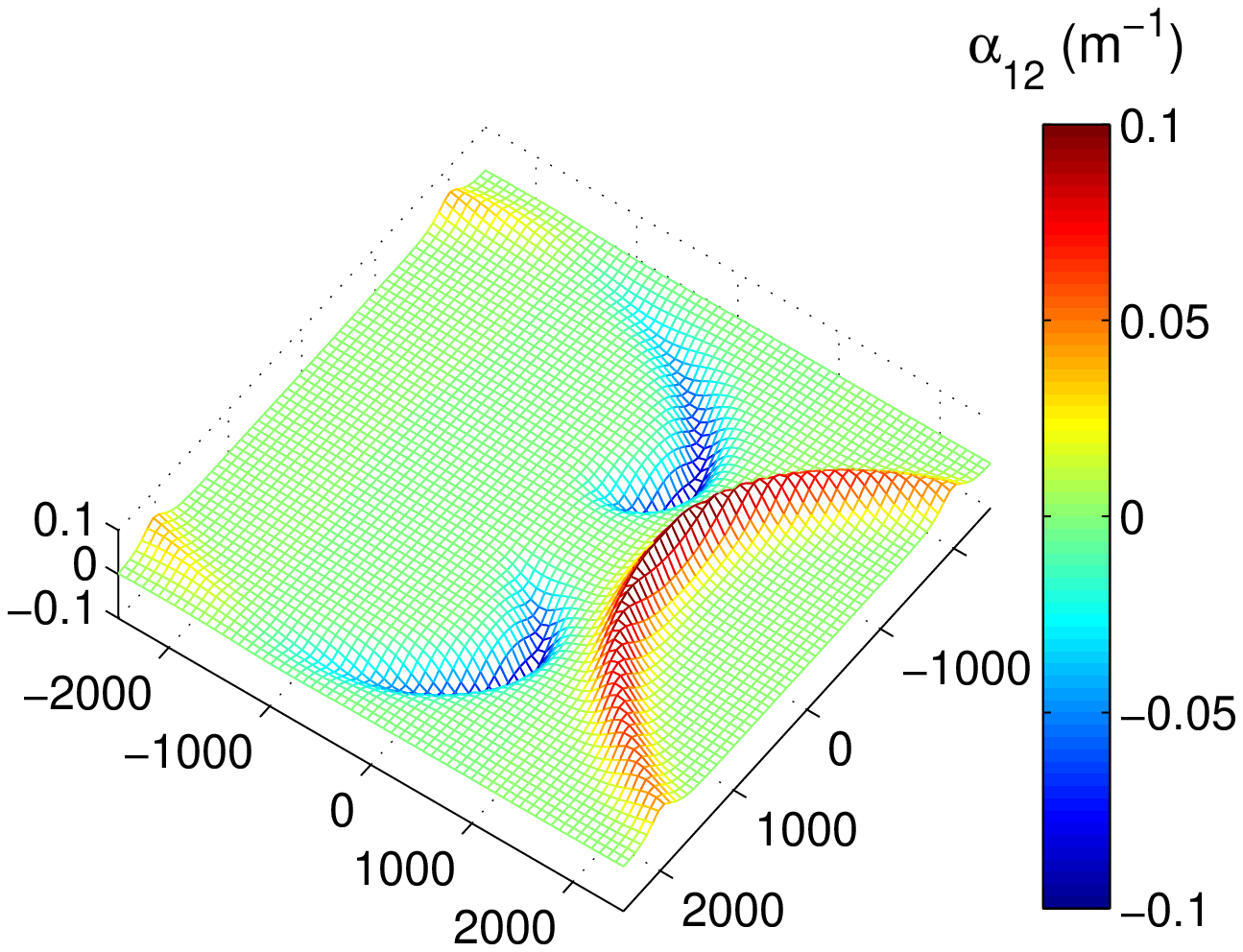} \\ (c) Step 140
  \end{center}
  \end{minipage}
  \begin{minipage}[b]{0.24\linewidth}
  \begin{center}
    \includegraphics[scale=0.25]{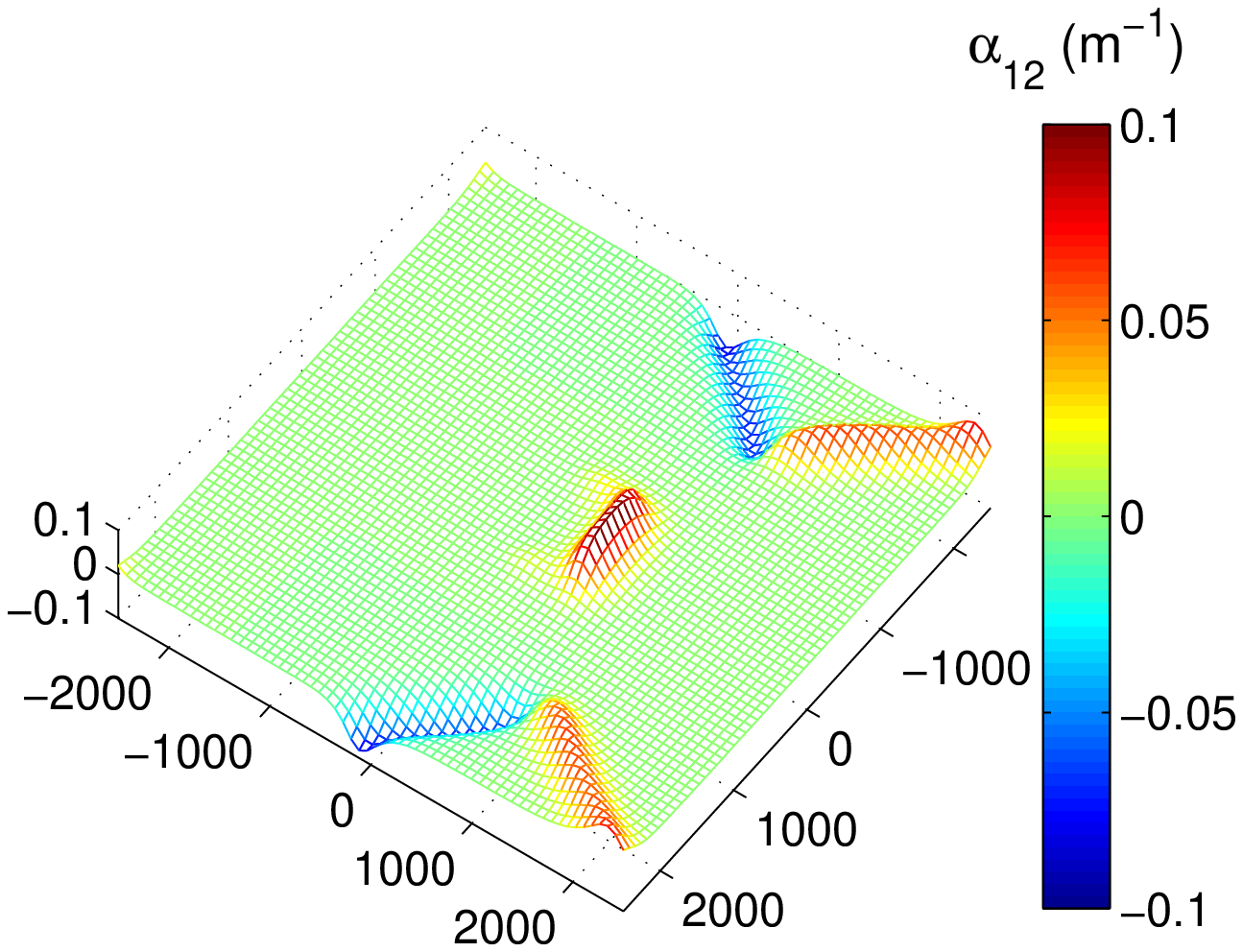} \\ (d) Step 150
  \end{center}
  \end{minipage}
  \caption{Example of the evolution of the dislocation density tensor field $\tensor{\alpha}(\vect{x^d})$ during a Frank-read source process simulated with the current DDD-FFT model. The initial source has a Burgers vector along the $x$-direction and is aligned along the $y$-direction. Top row: discrete dislocation structure, middle row: $\alpha_{11}^{\rm ns}$ component, bottom row: $\alpha_{12}^{\rm ns}$ component.}
  \label{fig:frsalpha}
\end{figure}

To illustrate potential routes to establish a connection between these models, consider as an example the simple, yet relevant process of the Frank-Read source mechanism. In Fig.~\ref{fig:frsalpha}, the bowing-out of a Frank-Read source under constant applied stress, simulated with the DDD-FFT framework, is presented. In this example, an initially straight, edge dislocation source has Burgers vector $\vect{b} = [b\,0\,0]$ and is aligned along the $y$-direction. As shown, the present DDD-FFT approach intrinsically captures the evolution of the dislocation tensor field $\tensor{\alpha}$ in response to the imposed loading. Despite its apparent triviality, the case of a Frank-Read source is enlightening as its evolution is not only governed by the imposed stress, but importantly involves the line-tension force. As a result, a simulation like that presented in Fig.~\ref{fig:frsalpha} is currently out of reach in the context of the FDM, highlighting the limitations of the approach discussed earlier. While a direct FDM solution that incorporates the effects of internal stresses in an explicit manner seems to be elusive at the present time, the DDD-FFT approach alternatively offers an interesting basis to compare, validate, and potentially inform phenomenological evolution laws to be incorporated in FDM models.

\subsection{Virtual X-ray diffraction} \label{sec:xrd}

In this section, we intend to show through a specific example that the full-field nature of the present approach offers some benefits that can be useful for other DDD-related applications. A relevant example is the emulation of virtual X-ray diffraction (XRD) patterns associated with discrete dislocation structures, which offers an interesting opportunity to compare features of the dislocation networks generated through simulations with experimental results.

As recently discussed in \cite{Bertin18b}, in which a non-singular computational method to compute virtual micro-Laue patterns was developed, the generation of such patterns from DDD configurations can be performed by ray-tracing and integrating the intensity of outgoing light rays reflected off crystallographic planes. To capture the elastic strain and rotation heterogeneities arising from the presence of dislocations, the method relies on the computation of the deformation gradient at a set of $N_d$ sampling points spanning the simulation volume \cite{Bertin18b}. As a result, the computation cost associated with such a procedure scales with $\mathcal{O}(N_{\rm img}^3 N_{\rm seg} N_d)$ for periodic dislocation structures, where $N_{\rm img}$ is the number of periodic images to be included in the calculation. While this cost surely remains tractable, it is nonetheless expensive; for instance, it was found that the total time required to calculate the deformation gradient field at $N_d=128^3$ sampling points for a configuration of $N_{\rm seg} = 15,000$ segments and using $N_{\rm img}=3$ periodic replica in each direction, amounts to approximately $30$ hours on a desktop CPU, or $40$ minutes with a GPU implementation \cite{Bertin18b}.

Alternatively, the present full-field approach directly provides the (long-range) distortion field $\tensor{U}^e$ (i.e. deformation gradient) arising from the dislocation structure (see \S\ref{sec:longrange}). As such, the total distortion field can be computed following a similar splitting as that introduced in Eq.~\eqref{eq:stresstot}, where the short-range contribution is calculated by summing the contributions of neighboring segments using the non-singular displacement gradient expression introduced in \cite{Bertin18b}. Since the long-range component is computed using the efficient FFT algorithm while a small number of neighbor segments $N_{\rm nei}$ must be typically included in the short-range calculation at each sampling point, the total computational cost is reduced to $\mathcal{O}(N_d\log N_d+ N_{\rm nei}N_d)$ when using the DDD-FFT framework. For instance, the cost associated with the evaluation of the total displacement gradient field is greatly reduced to below 4 minutes (on a desktop CPU) for the same configuration and sampling resolution as the example mentioned earlier, yielding a speed-up factor of about $500$. In addition, we note that using the DDD-FFT approach eliminates the issue of the conditional convergence that arises when evaluating the displacement gradient in periodic structures (see discussion in \cite{Bertin18b}), since the periodic nature of mechanical fields is intrinsically enforced in the spectral-based formulation.

\begin{figure}[t]
  \begin{minipage}[b]{0.5\linewidth}
  \begin{center}
    \includegraphics[scale=1.0]{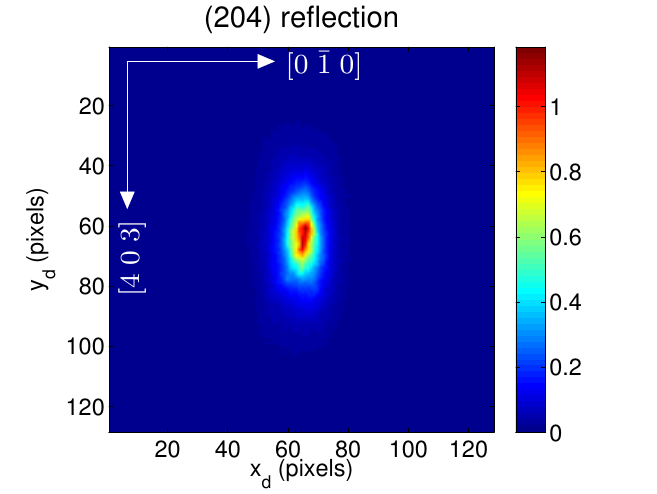} \\ (a)
  \end{center}
  \end{minipage}
  \begin{minipage}[b]{0.5\linewidth}
  \begin{center}
    \includegraphics[scale=0.4]{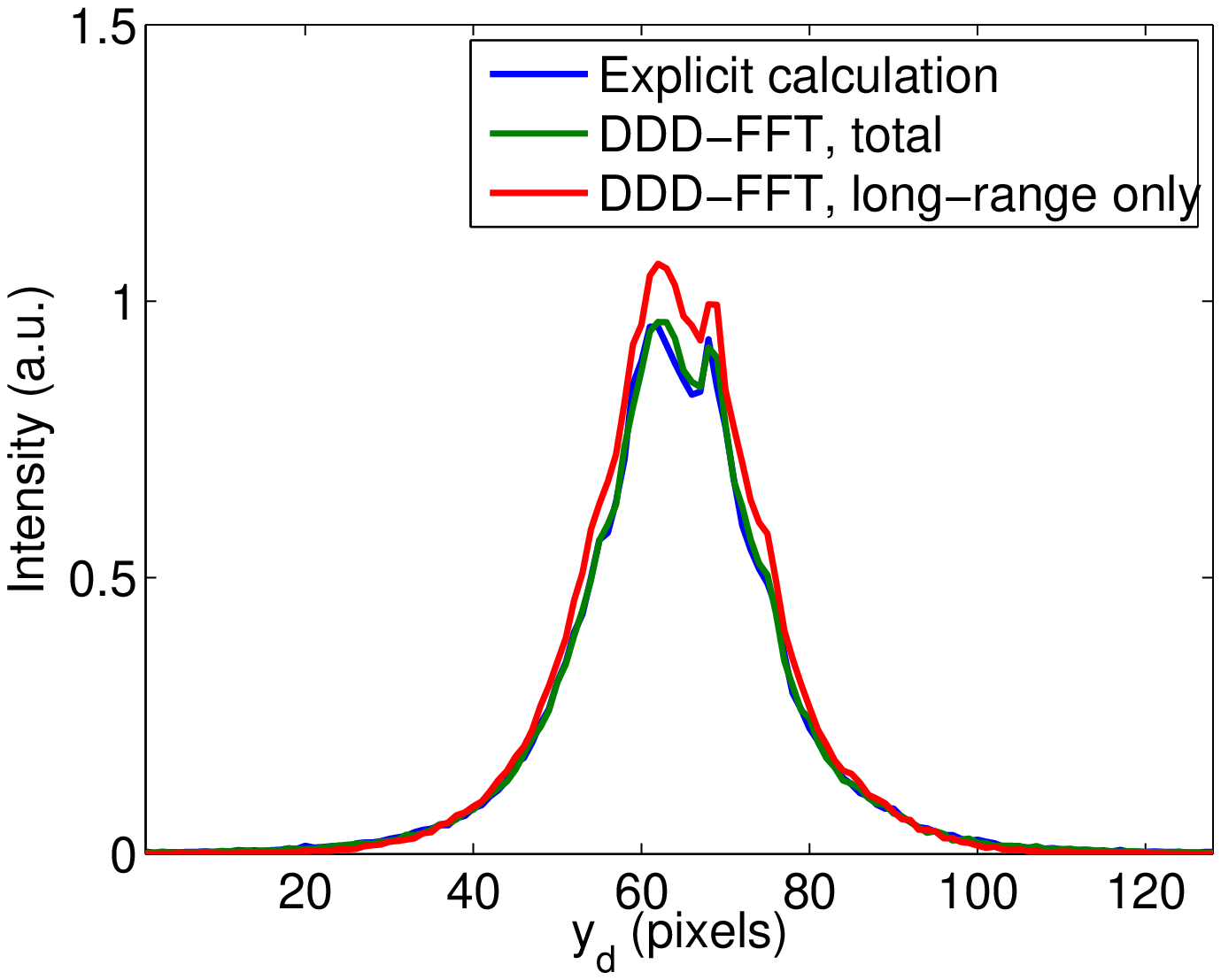} \\ (b)
  \end{center}
  \end{minipage}
  \caption{(a) Example of a $(204)$ micro-Laue virtual diffraction pattern obtained from the displacement gradient calculated with the DDD-FFT approach. The pattern is generated using the ray-tracing procedure detailed in \cite{Bertin18b}. (b) Peak intensity along a row of pixels passing trhough the center of the pattern in (a), when accounting for the total displacement gradient or the long-range contribution only.}
  \label{fig:virtualxrd}
\end{figure}

An example of the virtual $(204)$ XRD peak generated from the $(15 \mu m)^3$ periodic simulation reported in Fig.~\ref{fig:dddresults} is plotted in Fig.~\ref{fig:virtualxrd}(a) for a configuration taken at $\gamma = 0.5\%$ shear strain. As shown in Fig.~\ref{fig:virtualxrd}(b), the peak intensity distribution computed using the present spectral approach matches well that obtained using the explicit non-singular method developed in \cite{Bertin18b}. The slight difference between both peak intensities is believed to originate from periodic images effect; while $N_{\rm img}=3$ periodic replica were used in each direction in the explicit calculation, the spectral approach virtually accounts for an infinite number of images. Interestingly, it is further observed that the peak obtained when only considering the long-range contribution also leads to a rather good agreement, thereby highlighting the weak effect of short-range (core) contributions on the resulting patterns. This observation is consistent with results reported in \cite{Bertin18b}, where XRD patterns were found to exhibit little sensitivity to the choice of the physical core radius width $a_0$. Essentially, this result implies that a reasonable approximation of diffraction patterns can be obtained on the fly at virtually no cost during the course of a simulation when using the present DDD-FFT formulation.

\section{Discussion} \label{sec:discussion}

The improved DDD-FFT framework presented in this work relies in part on the dislocation density tensor to evaluate the stress fields associated with the presence of discrete dislocation lines. As a result, the first step of the approach consists in converting the set of discrete dislocation segments in the simulation volume into their continuum Nye's tensor representation, as presented in section \ref{sec:alphagrid}. We note here that a procedure, named D2C, was recently proposed in \cite{Sandfeld15} to perform this task. However, the procedure presented in our work differs from the D2C approach in two major aspects. In the D2C tool, dislocation lines are represented as spline segments and the calculation of the dislocation tensor involves a numerical quadrature integration, while a Gaussian smoothing kernel is used to spatially spread out dislocation lines. Alternatively, the method presented in section \ref{sec:alphagrid} for straight segments is fully analytical, and therefore allows to compute the dislocation density tensor very efficiently. In addition, our approach directly uses the non-singular distribution function as the spreading kernel for dislocations cores (see Eq.~\eqref{eq:alphaconv}). Note here that, in principle, the non-singular kernel could have been used directly in place of the CIC weighting function in Eqs.~\eqref{eq:IntSeg}-\eqref{eq:CIC3D}. However, no analytical expression could have been obtained for the line integral in Eq.~\eqref{eq:IntSeg}. Therefore our approach proceeds in two steps, whereby a ``singular'' dislocation tensor is first obtained, after which the non-singular kernel is applied using Eq.~\eqref{eq:alphaconv}.

In this regard, we note that, in the context of the remarkable FEM-based Discrete Continuous Model (DCM) introduced in \cite{Lemarchand01, Vattre14}, the use of the non-singular spreading function as a weighting function was proposed earlier to regularize the plastic strain distribution \cite{Liu09, Cui2015}. While a corrective short-range contribution was not included in these two earlier works, a similar splitting procedure to that introduced in Eq.~\eqref{eq:stresstot} was recently proposed in \cite{Jamond16}, again in the context of the DCM. However, here we point out several advantages that are offered by the use of the Nye's tensor compared to the eigenstrain-based DCM model, from which the original DDD-FFT formulation \cite{Bertin15} directly followed. In the eigenstrain-based formalism, the total stress field is evaluated from the plastic strain distribution arising from the motion of dislocations. Numerically, such an approach is not always optimal for several reasons. First, this is because the formulation is history-dependent, and therefore requires the knowledge of the accumulated plastic slip field at all times. For instance, it implies that a simulation cannot be restarted unless the corresponding history plastic is inputted. In addition, it necessarily follows that the approach is prone to numerical errors accumulation, with degrading accuracy during long simulation runs. Finally, the regularization of the plastic strain field requires an integration over the areas sheared by all dislocation segments \cite{Vattre14, Bertin15}. In comparison, relying on the dislocation density tensor eliminates the need for the plastic history knowledge, does not suffer from error accumulation, and solely involves line integrations to determine the total stress field, for which analytical expressions can be easily obtained (see section \ref{sec:alphagrid}). In addition, it potentially allows for an adaptive grid resolution scheme, whereby the mesh can be gradually refined as the dislocation density increases.

The DDD-FFT approach also offers several advantages compared to conventional DDD models. Among them, the FDM based spectral solver presented in section \ref{sec:longrange} provides an alternative solution to the Fast Multipole Method (FMM) \cite{LeSar02, Arsenlis07, Zhao10}, which is typically used to approximate long-range dislocation stress fields. In addition to providing an exact solution for the far-field contribution, the spectral solver is straightforward to implement, is computationally efficient thanks to the use of FFT algorithm, and is naturally suited for GPU computations. Furthermore, anisotropic elasticity comes at a same cost for the calculation of the long-range contribution, whereas it induces a cost factor of about 10 in the context of the FMM \cite{Yin12}. Lastly, it eliminates the issue of the conditional convergence of stress fields that arises in simulations subjected to periodic boundary conditions \cite{Cai03}.

Finally, compared to conventional DDD models, the full-field nature of the current DDD-FFT approach offers a convenient tool to establish direct connections with continuum theories of plasticity. In a similar intent, the D2C framework proposed in \cite{Sandfeld15} was introduced to allow for direct comparisons between Continuum Dislocation Dynamics (CDD) methods and DDD, and was used to examine the kinematic consistency of both approaches under prescribed velocity fields.  Similarly, it is expected that the current approach could be used to assess the validity of governing evolution laws in FDM.

\section{Conclusion} \label{sec:conclusion}

We developed a novel full-field DDD-FFT formulation that relies on the Nye's tensor to evaluate the stress and force fields resulting from the presence discrete dislocation segments. In order to benefit from a grid-based spectral approach while retaining submesh resolution details, the central idea of the method consists in taking advantage of the non-singular theory of dislocation by treating the dislocation core radius as a numerical parameter. With this, two different length-scales can be established and a rigorous splitting between short- and long-range stress contributions is introduced. While the long-range component is efficiently calculated from Nye's tensor using a FDM-based model, the short-range correction allowing to fully resolve submesh interactions is evaluated using non-singular analytical expressions typically used in conventional DDD implementations.

The accuracy and validity of the current DDD-FFT approach were assessed through different static and dynamics benchmarks. In all cases, the consistency of the approach and the validity of the stresses and forces splitting procedure was demonstrated. As an example, the method was further employed to perform large-scale work-hardening simulations in FCC single crystals.

While this new formulation allows to address several limitations that were inherent to the original DDD-FFT approach \cite{Bertin15}, it simultaneously retains all the advantages associated with its spectral nature. Specifically, it naturally incorporates the treatment of anisotropic elasticity, while providing a framework capable of dealing with heterogeneous elasticity \cite{Bertin18a}, thereby paving the way towards performing DDD simulations in multi-phases and polycrystalline materials. As for the original method, the formulation is particularly well suited for GPU-accelerated computation, for which an implementation was developed.

Finally, the use of the dislocation density tensor at the heart of the model allows to establish a direct connection with FDM approaches. Of particular interest, this approach can be regarded as a dynamical extension of FDM implementations, where the evolution of the dislocation tensor directly results from a DDD model in which the dislocation structure and evolving topology is explicitly resolved. Specifically, thanks to the fast analytical procedure developed to convert a set of discrete dislocation lines into their Nye's tensor representation, our approach allows to establish direct comparisons with FDM frameworks, thereby potentially allowing to inform and assess these models.

\appendix
\section{Analytical formulation for the CIC weighting function} \label{app:CIC}

In this appendix, a coordinate-independent analytical expression for line integral

\begin{equation}
I(\vect{x}^d,L^s) = \int_{\vect{x}^a}^{\vect{x}^b} S(\vect{x}^d-\vect{x}) \; dL(\vect{x})
\label{eq:IntSeg2}
\end{equation}

\noindent is presented when the integral is carried over a straight segment delimited by end points at positions $\vect{x}^a$ and $\vect{x}^b$, and the CIC weighting function defined in Eq. \eqref{eq:CIC3D} is used.

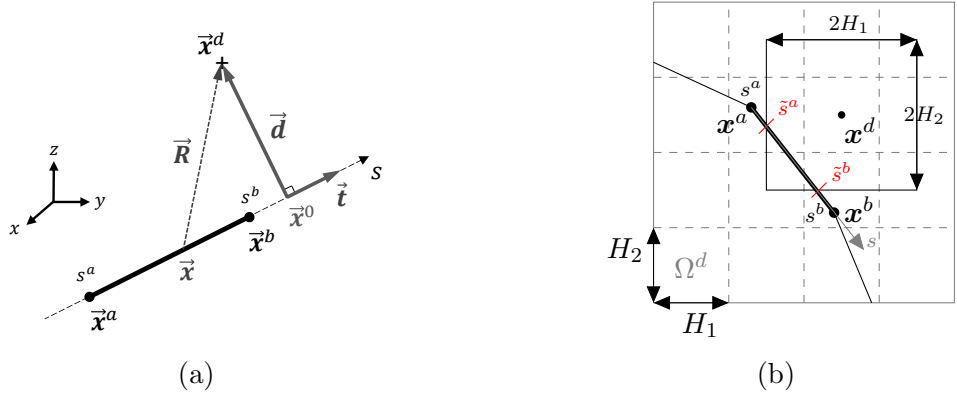
\begin{figure}[t]

\begin{minipage}[b]{0.5\linewidth}
\begin{center}
\includegraphics[scale=0.6]{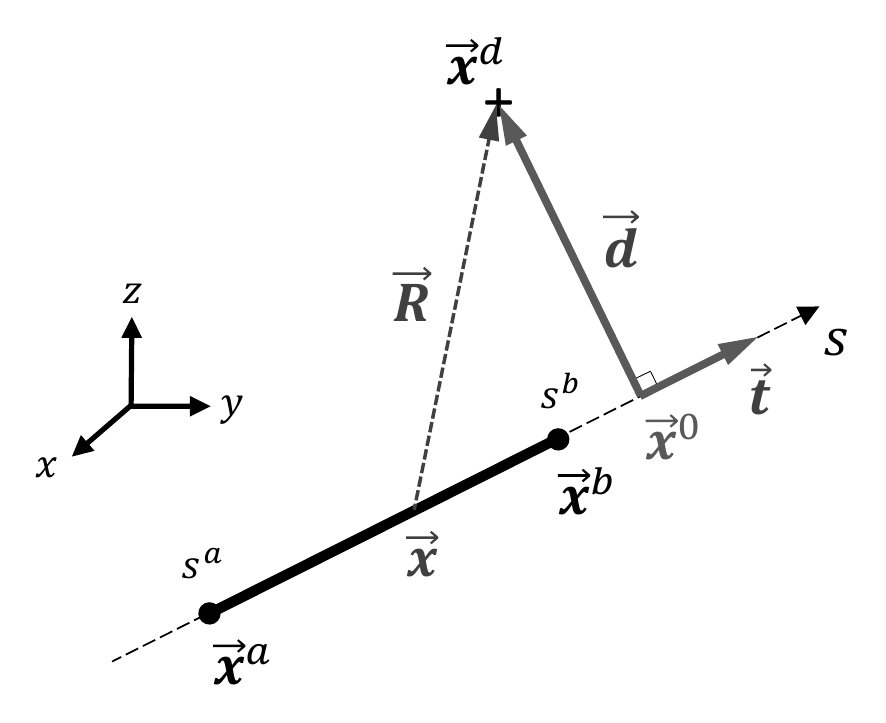} \\ (a)
\end{center}
\end{minipage} 
\begin{minipage}[b]{0.5\linewidth}
\begin{center}
\begin{tikzpicture}[scale=1.0, >=triangle 45]

\draw[gray] (0.0,0.0) rectangle (4.0,4.0);
\draw[gray,dashed] (1.0,0.0) -- (1.0,4.0);
\draw[gray,dashed] (2.0,0.0) -- (2.0,4.0);
\draw[gray,dashed] (3.0,0.0) -- (3.0,4.0);
\draw[gray,dashed] (0.0,1.0) -- (4.0,1.0);
\draw[gray,dashed] (0.0,2.0) -- (4.0,2.0);
\draw[gray,dashed] (0.0,3.0) -- (4.0,3.0);

\draw[<->] (0.0,0.0) -- (1.0,0.0) node[anchor=north east] {$H_1$};
\draw[<->] (0.0,0.0) -- (0.0,1.0) node[anchor=north east] {$H_2$};
\draw[gray] (0.5,0.5) node[] {$\Omega^d$};

\fill[black] (2.5,2.5) circle (0.05cm);
\draw (2.4,2.6) node[anchor=north west] {$\vect{x}^d$};
\draw[black] (1.5,1.5) rectangle (3.5,3.5);
\draw[<->] (1.5,3.5) -- (3.5,3.5);
\draw (2.6,3.7) node[] {\scriptsize $2H_1$};
\draw[<->] (3.5,3.5) -- (3.5,1.5);
\draw (3.6,2.5) node[] {\scriptsize $2H_2$};

\draw [] plot [smooth, tension=0] coordinates {(0.0,3.2) (1.3,2.6) (2.4,1.2) (2.9,0.0)};
\draw[ultra thick] (1.3,2.6) -- (2.4,1.2);
\fill[black] (1.3,2.6) circle (0.07cm);
\fill[black] (2.4,1.2) circle (0.07cm);
\draw (0.7,2.65) node[anchor=north west] {$\vect{x}^a$};
\draw (2.4,1.55) node[anchor=north west] {$\vect{x}^b$};

\draw (1.3,2.85) node[] {\scriptsize $s^a$};
\draw (2.15,1.2) node[] {\scriptsize $s^b$};

\draw[red] (1.4,2.25) -- (1.6,2.45);
\draw[red] (1.8,2.6) node[] {\scriptsize $\tilde{s}^a$};
\draw[red] (2.1,1.4) -- (2.3,1.6);
\draw[red] (2.5,1.75) node[] {\scriptsize $\tilde{s}^b$};

\draw[->,gray] (1.3,2.6) -- (2.8,0.7);
\draw[gray] (2.9,0.8) node[] {\scriptsize $s$};

\end{tikzpicture} \\ (b)
\end{center}
\end{minipage}

  \caption{(a) Parametrization used to describe the straight dislocation segment between end points $\vect{x}^a$ and $\vect{x}^b$. Point $\vect{x}^0$ is the orthogonal projection of vertex $\vect{x}^d$ on the dislocation line. The unit tangent $\vect{t}$ and vector $\vect{d} = \vect{x}^d - \vect{x}^0$ form an orthogonal basis, such that coordinate $\vect{x} = \vect{x}^0 + s\vect{t}$ spans segment $ab$ and the radius vector is defined by $\vect{R} = \vect{d} - s\vect{t}$ with $s \in \left[ s^a, s^b \right]$. (b) Cropping procedure to determine the parametric bounds $\tilde{s}^a$ and $\tilde{s}^b$ of the portion of segment $\vect{x}^a$-$\vect{x}^b$ to be included at voxel $\vect{x}^d$ in the CIC weighting scheme.}
  \label{fig:SegParam}
\end{figure}

For such a purpose, the following parametric representation of dislocation segment $L^s$ is adopted. As illustrated in Fig.~\ref{fig:SegParam}, let us denote $\vect{x}^0$ the orthogonal projection of vertex position $\vect{x}^d$ on the line supporting segment $L^s$, and $\vect{R}^d = \vect{x}^d-\vect{x}$ the distance vector linking the grid point $\vect{x}^d$ to the coordinate $\vect{x}$ spanning segment $s$. As such, the constant vector $\vect{d}$ linking grid point $\vect{x}^d$ to its projection $\vect{x}_0$ is given by:

\begin{gather}
\vect{d} = \vect{x}^d -\vect{x}^0 \quad \text{with} \quad \vect{d} \cdot \vect{t} = 0 \\
\text{such that} \quad \vect{d} = \vect{R}^d -(\vect{R}^d\cdot\vect{t})\vect{t}
\end{gather}

\noindent where $d=\|\vect{d}\|$ is the distance of the grid vertex $\vect{x}_d$ to the supporting line of segment $L^s$, and $\vect{t}$ is its unit tangent vector defined as:

\begin{equation}
\vect{t} = \frac{\vect{x}^b-\vect{x}^a}{\|\vect{x}^b-\vect{x}^a\|}
\end{equation}

\noindent With this, the segment line $L^s$ can be conveniently described with the following parametric representation:

\begin{equation}
\vect{x} = \vect{x}^0 + s\vect{t}, \quad s \in [s^a,s^b]
\end{equation}

\noindent where the bounds $s^a$ and $s^b$ for parameter $s$ are given by:

\begin{gather}
s^a = (\vect{x}^a-\vect{x}^0) \cdot \vect{t} \\
s^b = (\vect{x}^b-\vect{x}^0) \cdot \vect{t}
\end{gather}

\noindent With this setting, vectors $\vect{t}$ and $\vect{d}$ form an orthogonal basis such that the radius vector $\vect{R}^d$ can be expressed as:

\begin{equation}
\vect{R}^d = \vect{x}^d-\vect{x} = \vect{d}-s\vect{t}
\end{equation}

\noindent such that the perpendicular distances to be evaluated in the weight function in Eq.~\eqref{eq:CIC3D} can be conveniently calculated as: 

\begin{equation}
|x^d_i-x_i| = \| \vect{R}^d \cdot \vect{e}_i \| = \| \vect{R}^d_i \| \\
\end{equation}

\noindent where $\|\vect{R}^d_i\|$ denotes the norm of the projection of the distance vector $\vect{R}^d$ along the $i$-axis associated with basis vector $\vect{e}_i$. The norm of the distance vectors are given by:

\begin{equation} \label{eq:RadiusProj}
\|\vect{R}^d_i\| = \sqrt{\vect{R}^d_i \cdot \vect{R}^d_i} = \sqrt{\left( d_i-st_i \right)^2}
\end{equation}

\noindent Using the parametrization introduced in the above and the definition of the CIC weight function in Eq.~\eqref{eq:CIC3D}, line integral in Eq.~\eqref{eq:IntSeg2} is expressed as:

\begin{equation}
I(\vect{x}^d,L^s) = \int_{\tilde{s}^a}^{\tilde{s}^b} \left( 1 - \frac{\|\vect{R}^d_1\|}{H_1} \right) \left( 1 - \frac{\|\vect{R}^d_2\|}{H_2} \right) \left( 1 - \frac{\|\vect{R}^d_3\|}{H_3} \right) ds
\label{eq:IntSeg3}
\end{equation}

\noindent where $\tilde{s}^a$ and $\tilde{s}^b$ are the modified parametric bounds of segment $L^s$ to ensure that conditions $\|\vect{R}^d_i\| = |x_i^d-x_i| < H_i$, $\forall i \in \{ 1,2,3\}$ are satisfied on the integration domain. The modified bounds effectively correspond to the coordinates of the intersection of segment $L^s$ with the elementary weighting volume associated with each voxel $d$ of the grid $\Omega^d$. As illustrated in Fig.~\ref{fig:SegParam}(b), coordinates $\tilde{s}^a$ and $\tilde{s}^b$ can be conveniently determined by cropping segment $L^s$ with a bounding box of size $2H_1 \times 2H_2 \times 2H_3$ centered at voxel position $\vect{x}^d$.

Combining Eqs. \eqref{eq:RadiusProj} and \eqref{eq:IntSeg3}, the line integral can be written as:

\begin{equation}
I(\vect{x}^d,L^s) = ( \tilde{s}^b-\tilde{s}^a ) - A_1 - A_2 - A_3 + B_{12} + B_{13} + B_{23} - C_{123}
\label{eq:IntSeg4}
\end{equation}

\noindent where quantities $A_i$, $B_{ij}$ and $C_{ijk}$ are given by:

\begin{align} \label{eq:IntSeg5}
A_i &= \int_{\tilde{s}^{\alpha}}^{\tilde{s}^{\beta}} \frac{\sqrt{(d_i-st_i)^2}}{H_i} ds \nonumber \\
B_{ij} &= \int_{\tilde{s}^{\alpha}}^{\tilde{s}^{\beta}} \frac{\sqrt{(d_i-st_i)^2(d_j-st_j)^2}}{H_i H_j} ds \nonumber \\
C_{ijk} &= \int_{\tilde{s}^{\alpha}}^{\tilde{s}^{\beta}} \frac{\sqrt{(d_i-st_i)^2(d_j-st_j)^2(d_k-st_k)^2}}{H_i H_j H_k} ds
\end{align}

\noindent Note that care must be taken in evaluating analytically the quantities expressed in Eqs. \eqref{eq:IntSeg5}. This is because the sign of $\prod_i^n (d_i-st_i)$ (where $n = 1, 2$ or $3$ for quantities $A_i$, $B_{ij}$ and $C_{ijk}$, respectively) is likely to change over the interval $[\tilde{s}^{\alpha},\tilde{s}^{\beta}]$. To ensure positiveness of the $\ell^2$-norm during integration, the domain must first be divided into subsequent portions $[a,b] \subset [\tilde{s}^{\alpha},\tilde{s}^{\beta}]$ on which the sign of $\prod_i^n (d_i-st_i)$ does not change. On each such portions, the evaluation of line integrals defined in Eqs. \eqref{eq:IntSeg5} is then performed using the following analytical expressions:

\begin{align}
A_i &= \int_{a}^{b} \frac{\sqrt{(d_i-st_i)^2}}{H_i} ds \nonumber \\
&= \frac{1}{H_i} \abs{ \int_{a}^{b} \left( d_i-st_i \right) ds } = \frac{1}{H_i} \abs{ \left[ d_i s -\frac{1}{2} t_i s^2\right]_{a}^{b} }
\end{align}

\begin{align}
B_{ij} &= \int_{a}^{b} \frac{\sqrt{(d_i-st_i)^2(d_j-st_j)^2}}{H_i H_j} ds \nonumber \\
&= \frac{1}{H_i H_j} \abs{ \int_{a}^{b} \left( d_i-st_i \right) \left( d_j-st_j \right) ds } \nonumber \\
&= \frac{1}{H_i H_j} \abs{ \left[ d_i d_j s -\frac{1}{2} (d_i t_j + d_j t_i) s^2 + \frac{1}{3} t_i t_j s^3 \right]_{a}^{b} }
\end{align}

\begin{align}
C_{ijk} &= \int_{a}^{b} \frac{\sqrt{(d_i-st_i)^2(d_j-st_j)^2(d_k-st_k)^2}}{H_i H_j H_k} ds \nonumber \\
&= \frac{1}{H_i H_j H_k} \abs{ \int_{a}^{b} \left( d_i-st_i \right) \left( d_j-st_j \right) \left( d_k-st_k \right) ds } \nonumber \\
&= \frac{1}{H_i H_j H_k} \left| \left[ d_i d_j d_k s -\frac{1}{2} (t_i d_j d_k + d_i t_j d_k + d_i d_j t_k) s^2 \right. \right. \nonumber \\
&\phantom{\frac{1}{H_i H_j H_k}\qquad} \left. \left. + \frac{1}{3} (d_i t_j t_k + t_i d_j t_k + t_i t_j d_k) s^3 - \frac{1}{4} t_i t_j t_k s^4 \right]_{a}^{b} \right|
\end{align}

\section{Error quantification in the stress splitting procedure} \label{app:splittingerror}

In the non-singular dislocation theory, it can be shown that the stress field of an elementary dislocation segment behaves as $R^{-2}$ at large $R$ \cite{Cai06}:

\begin{equation}
\sigma_{ij}^{\rm ns}(\vect{R},a) \propto \frac{1}{R_a^2} = \frac{1}{R^2+a^2}
\end{equation}

\noindent where $R = \normv{\vect{R}}$ is the distance measured from the dislocation segment, and $R_a = \sqrt{R^2+a^2}$ is the non-singular radius vector associated with core radius $a$. Taking a series expansion, it can be shown that the difference between the stress fields evaluated for different values $a_1$ and $a_2$ of the core radius decays as:

\begin{equation}
\abs{\sigma_{ij}^{\rm ns}(\vect{R},a_1) - \sigma_{ij}^{\rm ns}(\vect{R},a_2)} \propto \frac{a_2^2 - a_1^2}{R^4} + O(\frac{1}{R^6}) \quad \text{as } R \rightarrow \infty
\end{equation}

\noindent where $a_2 > a_1$. As a result, the relative difference with respect to the stress obtained at $a_1$ behaves as $1/R^2$. A value for the critical $r_c$ radius beyond which the maximum relative error becomes smaller than a prescribed tolerance $\epsilon_{\rm max} > 0$ is simply obtained as:

\begin{equation}
r_c(a_1,a_2) = \sqrt{\frac{a_2^2 - a_1^2}{\epsilon_{\rm max}}}
\end{equation}

\begin{figure}
  \begin{center}
    \includegraphics[width=7.5cm]{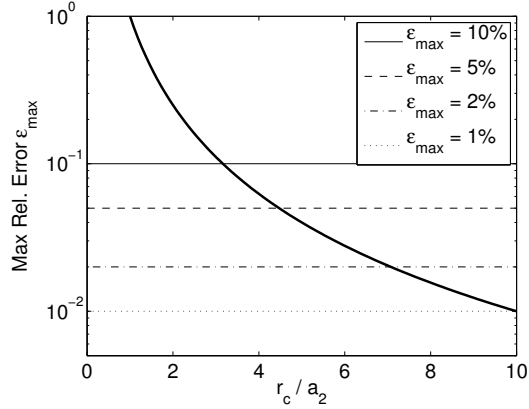}
    \vspace*{-0.5cm}
  \end{center}
  \caption{Maximum relative splitting error $\epsilon_{\rm max}$ as a function of the critical radius $r_c$, for $a_1 = 1b$ and $a_2 = 1000b$.}
  \label{fig:rc_a2_err}
\end{figure}

\noindent An example of the maximum splitting error introduced as a function of the value of the critical radius $r_c$ is reported in Fig.~\ref{fig:rc_a2_err} for typical core radius values of $a_1 = 1b$ and $a_2 = 1000b$.

\bibliographystyle{ieeetr}
\bibliography{ddd_fft.bib}

\end{document}